\documentclass[aps,pre,twocolumn,floatfix,10pt]{revtex4-1}
\usepackage{graphicx}
\usepackage{dcolumn}
\usepackage{bm}
\usepackage{latexsym}
\usepackage{xcolor}
\usepackage{amsmath}
\usepackage{amsmath}
\newcommand{\RN}[1]{%
  \textup{\uppercase\expandafter{\romannumeral#1}}%
}
\begin{document}
\title{{\bf 
Interference of two co-directional exclusion processes\\ in the presence of a static bottleneck: 
a biologically motivated model}}
\author{Bhavya Mishra}
\affiliation{Department of Physics, Indian Institute of Technology Kanpur, 208016, India}
\author{Debashish Chowdhury{\footnote{Email: debch@iitk.ac.in}}}
\affiliation{Department of Physics, Indian Institute of Technology Kanpur, 208016, India}

\begin{abstract}
We develope a two-species exclusion process with a distinct pair of entry and exit sites for each 
species of rigid rods. The relatively slower forward stepping of the rods in an extended bottleneck 
region, located in between the two entry sites, controls the extent of interference of the 
co-directional flow of the two species of rods. The relative positions of the sites of entry of 
the two species of rods with respect to the location of the bottleneck are motivated by a biological 
phenomenon. However, the primary focus of the study here is to explore the effects of the 
interference of the flow of the two species of rods on their spatio-temporal organization and the 
regulations of this interference by the extended bottleneck. By a combination 
of mean-field theory and computer simulation we calculate the flux of both species of rods and their 
density profiles as  well as the composite phase diagrams of the system. If the bottleneck is 
sufficiently stringent some of the phases become practically unrealizable although not ruled out on 
the basis of any fundamental physical principle. Moreover the extent of suppression of flow of the 
downstream entrants by the flow of the upstream entrants can also be regulated by the strength of the 
bottleneck. We speculate on the possible implications of the results in the context of the biological 
phenomenon that motivated the formulation of the theoretical model.

\end{abstract}

\maketitle
\section{Introduction}
Non-equilibrium steady states (NESS) of systems of interacting driven 
particles are of current interest in statistical physics 
\cite{evans05,blythe07}. 
Totally asymmetric simple exclusion process (TASEP)
\cite{schutz01,derrida98,mallick15} 
is a paradigmatic model of interacting self-propelled particles
\cite{schutz01,derrida98,mallick15,mukamel00,blythe07}.
In this model a fraction of the sites on a one-dimensional lattice are 
occupied by particles that can hop forward probabilistically, at a given 
rate (i.e., with a probability per unit time), if and only 
if its target site is empty. Various adaptations and extensions of TASEP, 
including multi-species exclusion processes 
\cite{chowdhury00,schad10,chowdhury05,chou11,chowdhury13a,rolland15,macdonald68,macdonald69,tripathi08,klumpp08,klumpp11,sahoo11,ohta11,wang14,schutz03,kunwar06,john04,lin11,lakatos03,shaw03,shaw04a,shaw04b,chou03,chou04,zia11,chou99,levine04,liu10,brackley12,greulich12,basu07,gccr09,ciandrini10,kuan16,chowdhury08,oriola15,sugden07,evans11,chai09,ebbinghaus09,ebbinghaus10,muhuri10,neri11,neri13a,neri13b,curatolo16,klein16,parmeggiani04,graf17},
have been used to model vehicular traffic (see \cite{chowdhury00,schad10} 
for reviews) and traffic-like collective phenomena in biological systems 
(see ref.\cite{chowdhury05,chou11,chowdhury13a,rolland15} for reviews). 
Under open boundary conditions, the rates of entry and exit of the particles 
are also specified. In analogy with vehicular traffic, the points of entry 
and exit of the particles are often also referred to as the ON- and OFF- ramps, 
respectively. By convention adopted  in this paper, all the particles 
move from left to right.

Special sites  from which the particles can hop forward at a relatively slower 
rate are usually referred to as defect sites. An isolated slow site is a ``point-like'' 
defect whereas a continuous string of slow sites constitute a ``line-like'' (or 
``extended'') 
defect. Each of these defects, irrespective of its size, creates a bottleneck 
where the flow of particles slows down. The effects of a single bottleneck 
as well as those of randomly distributed bottlenecks on the spatio-temporal 
organization of the particles, particularly the flow in the NESS, have been 
investigated over the last two decades 
\cite{dong12,greulich08a,barma06,kolimeisky98,pierobon06,harris04,enaud04,schmidt15,
embley09,foulaadvand08,greulich08b,sarkar14,banerjee15,
dhiman16,daga17,cirillo16,cook13,dhiman17,dong07a,foulaadvand07,
goldstein98,ha03,hinrichsen97,janowsky92,janowsky94,juhasz05,juhasz06,kolwankar00,krug91,
krug00,laktos05,mallick96,schutz93,tripathy97,tripathy98,pottmeier02,dong07b}.
In all those models with bottlenecks only a single ON-, OFF-ramp pair was 
considered. 

In this paper we develop a biologically motivated two-species exclusion 
process with a distinct pair of ON-, OFF-ramps for each species, where an 
extended defect lies in between the two ON ramps. More specifically, 
the ON-ramp of one of the two species of particles is located immediately 
downstream from the right edge of a single extended bottleneck while the 
other ON ramp lies far upstream from the left edge of that defect. 

Although the 
relative positions of the two ON-ramps with respect to the extended bottleneck 
in this model is motivated by a specific biological phenomenon, it is not intended to account 
for experimental data. Instead, the model focuses on the physics of collective 
spatio-temporal organization of the two species of particles. By influencing 
the flow of the particles that enter through the upstream ON-ramp, the bottleneck 
can control the extent of interference of the flow the two species in the downstream 
region. For this model, by a combination of mean-field theory and computer simulations, 
we draw the phase diagrams that displays an unprecedented richness. We also 
demonstrate a switch-like regulation of flow of one species of rods by a sufficiently 
high flow of the other.

\section{Model}

In the first two subsections we present brief summaries of TASEP with hard rods and 
the biological phenomenon of Internal Ribosome Entry Site (IRES) that motivates 
the model introduced in this paper.

\subsection{Brief summary of TASEP with hard rods and models of ribosome traffic}
In a TASEP under open boundary conditions equispaced sites on a 
one-dimensional lattice are labeled by the integer index 
$j=1,2,...,L+{\ell}-1$. In the case TASEP with a single species 
of hard rods, each of length ${\ell}$ (in the units of lattice 
spacing), successive ${\ell}$ sites are {\it covered} simultaneously 
by each rod. The {\it position} of each rod on the lattice is denoted by 
the lattice site covered by its leftmost edge. In our terminology 
used throughout this paper, a site $j$ is said to be {\it occupied} 
by a rod if $j$ denotes its position. Thus, at any arbitrary instant 
of time, if a site $j$ is {\it occupied} by a rod then all the sites 
$j,j+1,j+2,...,j+{\ell}-1$ are simultaneously {\it covered} by the 
same rod. Thus, out of the $L+{\ell}-1$ lattice sites only the $L$ 
sites $1,2,...,L$ can be {\it occupied} by the rods; the remaining 
${\ell}-1$ sites can be covered by a rod if it is occupies the site $L$.

Mutual exclusion of the rods is ensured by imposing the condition 
that no lattice site can be covered by more than one rod simultaneously. 
At any arbitrary instant of time the {\it number density} of the rods on 
the lattice is defined by 
\begin{equation}
\rho = N/L
\label{eq-numden} 
\end{equation}
where $N$ is the total number of rods occupying lattice sites at that 
instant of time. Since each rod simultaneously covers ${\ell}$ sites 
on the lattice and since none of the sites can be covered by more than 
one rod simultaneously the {\it coverage density} is 
given by
\begin{equation}
\rho_{c}=\rho\ell
\label{o-t-c}
\end{equation}
where the corresponding number density $\rho$ is obtained from 
eq.(\ref{eq-numden}). In other words, coverage density is the true 
measure of the total fraction of lattice sites occluded by the 
rods while only the remaining fraction remains empty.

Under open boundary conditions, a rod can enter the lattice only through 
the site $i=1$. If simultaneously all the first 
$\ell$ sites of the lattice (i.e., the sites $1,2,...,{\ell}$) are not 
covered by any existing rod a new rod can enter the system and occupy the 
site $i=1$ (and also cover the sites $2,...,{\ell}$ simultaneously); 
the rate of this event is $\alpha$. Similarly, a rod can exit from the 
lattice, with rate $\beta$, if it is occupying the site $i=L$ (and covering 
the remaining sites $i=L+1,L+2 \dots L+\ell-1$). A rod that occupies 
any site $i$ other than $i=L$ can hop forward, with step size unity 
(measured also in the unit of lattice spacing) and jump rate $p$, only if 
the site $i+{\ell}$ is empty. \\

\begin{figure}[t] 
\begin{center}
\includegraphics[width=1.0\columnwidth]{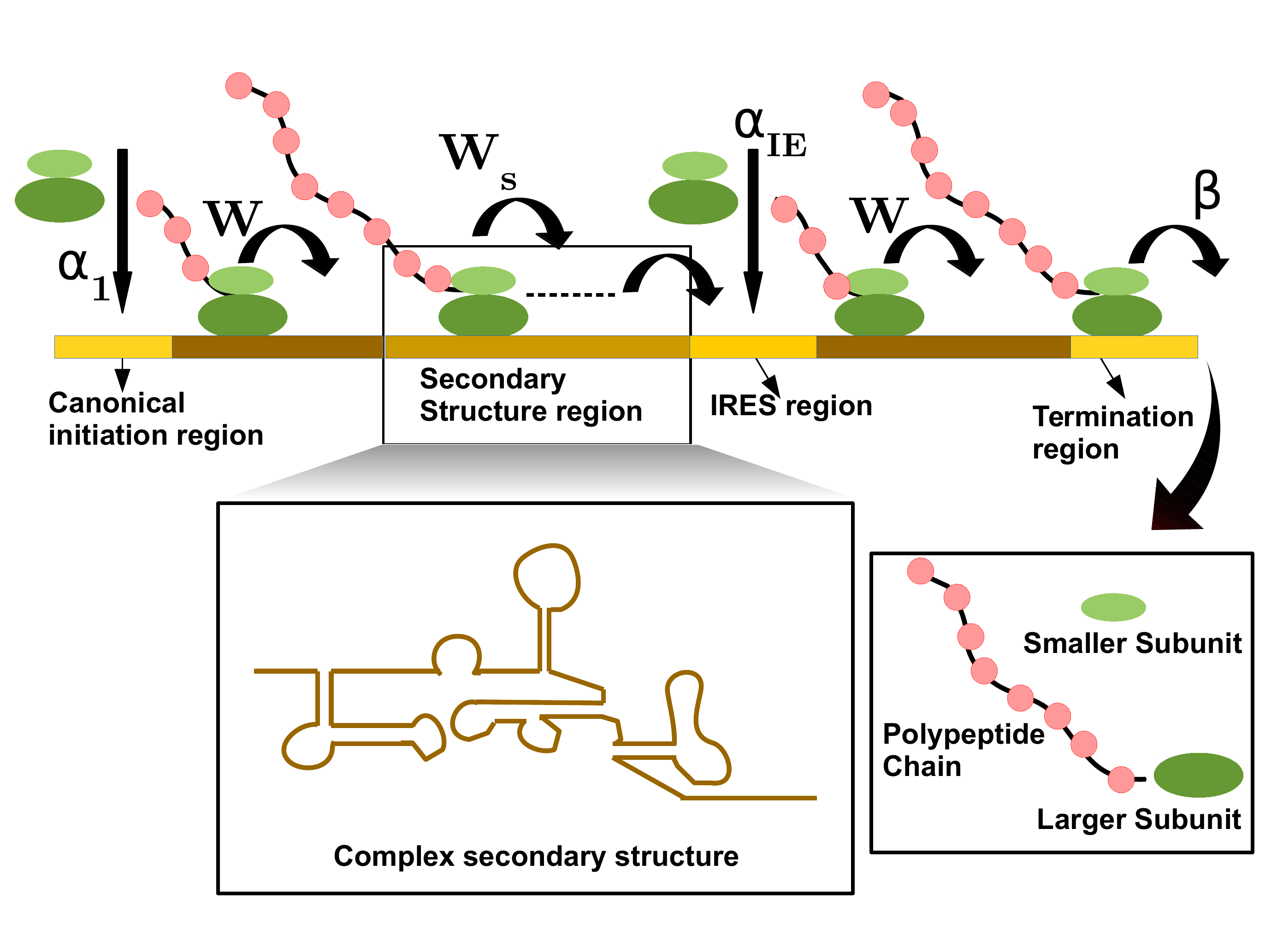} 
\end{center} 
\caption{(Color online) A cartoon depicting the phenomenon of unconventional 
translation initiated through IRES. The relative positions of the sites 
of canonical initiation and IRES with respect to that of the secondary 
structure of the mRNA track, along with the rates of the various kinetic 
processes, are shown schematically (see the text for details).} 
\label{ires_cartoon}
\end{figure}

\begin{figure}[t] 
\begin{center}
\includegraphics[width=1.0\columnwidth]{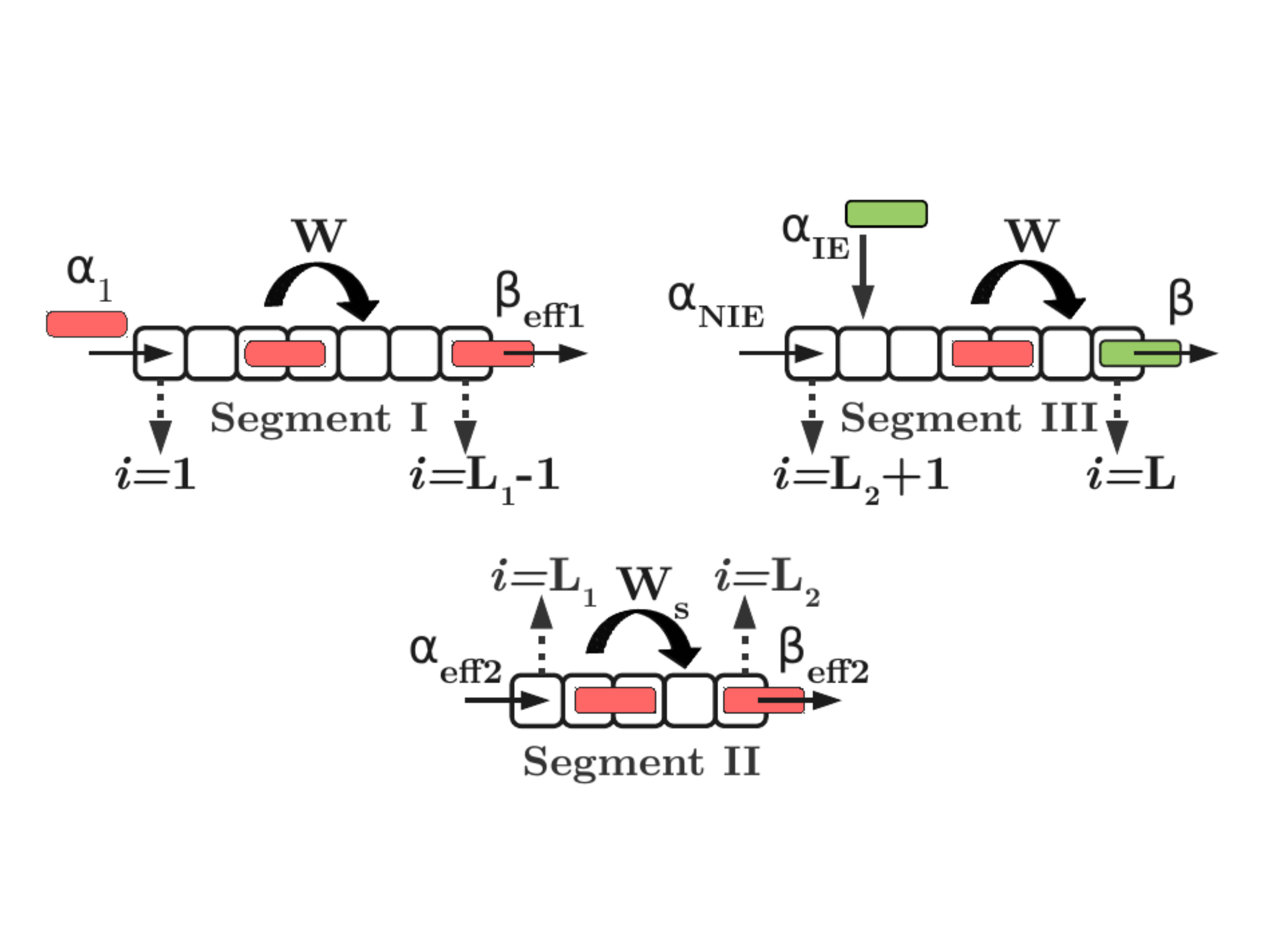} 
\end{center} 
\caption{(Color online) A schematic description of the two-species exclusion model, 
intended to capture some of the key features of Fig.\ref{ires_cartoon}. 
The whole lattice is divided into three segments. Segments \RN{1} and 
\RN{3} are regions where the rate of forward hopping of a rod is $W$, 
irrespective of the site and identity of the rod. The mid-segment \RN{2}, 
that mimics the secondary structure of mRNA in Fig\ref{ires_cartoon}, 
is essentially an extended bottleneck because the forward jump rate $W_s$ 
of the rods in this segment is less than $W$. Sites $i=1$ and 
$i=i_s=L_2+n\ell+m$ are the sites of entry for the $rod_1$ (with rate 
$\alpha_{1}$ in segment \RN{1}) and for $rod_2$ ( with rate $\alpha_{IE}$ 
in segment \RN{3}), respectively. Thus, segments \RN{1} and \RN{2} 
are populated by only $rod_1$ whereas a mixed population of $rod_1$ and 
$rod_2$ exists in segment \RN{3}. $\alpha_{eff2}$ and $\alpha_{eff3}$ 
are effective rates of entry of the rods into the segments \RN{2} and \RN{3}, 
respectively; $\alpha_{eff3}$ gets contributions from $\alpha_{IE}$ as 
well as $\alpha_{NIE}$, the latter being the effective rate of entry of 
$rod_1$ from segment \RN{2} into segment \RN{3}.  Similarly, $\beta_{eff1}$ 
and $\beta_{eff2}$ are the effective rates of exit of the $rod_1$ from 
segments \RN{1} and \RN{2}, respectively. For simplicity, we assume both 
species of rods to exit from the lattice, with the same rate $\beta$, 
from the same site $i=L$, irrespective of their identity. In this figure 
the rod length is taken as $\ell=2$ just for the purpose of illustration; 
in actual numerical calculations rod size has been taken as $\ell=10$ 
which captures the size of a ribosome more realistically.} 
\label{ires_model}
\end{figure}
$ P(i,t)$ denotes the probability of finding a rod at site $ i $ at time $ t $.\\
We define $ p(\underline{i} \mid i+\ell) $ as the conditional probability of finding site $ i+\ell $ empty, if site $ i $ is given to be occupied. It is straightforward to show that \cite{basu07,gccr09}
\begin{equation}
p(\underline{i} \mid i+\ell)=\dfrac{1-\rho \ell}{1+\rho-\rho \ell},
\label{conditional_1}
\end{equation}
where, $ \rho $ is the occupational density of the system.\\
The master equations corresponding to the above explained dynamics are given below,\\
\textbf{a:} At site $ i=1 $,
\begin{equation}
\dfrac{dP(i,t)}{dt}=\alpha P(\underbrace{0,....,0})-P(i,t)p(\underline{i} \mid i+\ell)p, 
\label{site1}
\end{equation}
where, $ P(\underbrace{0,....,0}) $ is the probability that all sites from $ i=1 $ to $ i=\ell $ are empty.\\
\textbf{b:} At site $ i=L $,
\begin{equation}
\dfrac{dP(i,t)}{dt} =P(i-1,t)p -P(i,t)\beta,
\label{siteN}
\end{equation}
\textbf{c:} At all remaining sites,
\begin{equation}
\begin{split}
\dfrac{dP(i,t)}{dt} & =P(i-1,t)p(\underline{i-1} \mid i+\ell-1)p \\
& -P(i,t)p(\underline{i} \mid i+\ell)p. 
\end{split}
\label{rem_sites}
\end{equation}
Under steady state condition, flux $ (J) $ in a uniform system is given by,\\
\begin{align}
J=p\dfrac{\rho(1-\rho\ell)}{(1+\rho-\rho\ell)}
\label{flux_a}
\end{align}
By taking, $ \dfrac{dJ}{d\rho}=0 $, we get,\\
\begin{eqnarray}
J_{MC}=
        \begin{cases}
            \dfrac{1}{(\sqrt{\ell}+1)^2}& \text{if $p=1$}, \\
             \dfrac{p}{(\sqrt{\ell}+1)^2}& \text{if $p<1$} ,
         \end{cases}
         ~~~~~~~~(a)
\nonumber \\
\nonumber\\
\rho_c^{MC}=
        \begin{cases}
            \dfrac{\sqrt{\ell}}{(\sqrt{\ell}+1)}& \text{if $p=1$} ,\\
            \dfrac{\sqrt{\ell}}{(\sqrt{\ell}+1)} & \text{if $p<1$}. 
            \label{mcp_2}
         \end{cases}
          ~~~~~~~~(b)
\end{eqnarray}
If initiation is the rate-limiting step (i.e., the rate $\alpha$ is the 
smallest of the three rates), then the system exhibits the low density 
(LD) phase. The steady state flux and the coverage density in 
the LD phase are given by 
\begin{eqnarray}
J_{LD}&=&
        \begin{cases}
            \dfrac{\alpha(1-\alpha)}{1+\alpha(\ell-1)}& \text{if $p=1$}, \\
             \dfrac{\alpha(p-\alpha)}{p+\alpha(\ell-1)}& \text{if $p<1$},
         \end{cases}
      ~~~~~~~~(a)
\nonumber \\
\nonumber\\
\rho_c^{LD}&=&
        \begin{cases}
            \dfrac{\alpha\ell}{1+\alpha(\ell-1)}& \text{if $p=1$}, \\
             \dfrac{\alpha\ell}{p+\alpha(\ell-1)}& \text{if $p<1$}.
          \end{cases}
          ~~~~~~~~~(b)
          \label{ldp_2}
\end{eqnarray}
\\
Similarly, if termination is the rate-limiting (i.e., slowest) step, the system would be in the high density (HD) phase. The steady state flux and the coverage density in the HD phase are given by
\begin{eqnarray}
J_{HD}&=&
        \begin{cases}
            \dfrac{\beta(1-\beta)}{1+\beta(\ell-1)}& \text{if $p=1$}, \\
             \dfrac{\beta(p-\beta)}{p+\beta(\ell-1)}& \text{if $p<1$},
         \end{cases}
      ~~~~~~~(a)
\nonumber \\
\nonumber\\
\rho_c^{HD}&=&
        \begin{cases}
            1-\beta& \text{if $p=1$}, \\
            \dfrac{(p-\beta)}{p} & \text{if $p<1$}. 
          \end{cases}
          ~~~~~~~~~~~~~~(b)
          \label{hdp_2}
\end{eqnarray}
By comparing the fluxes (\ref{mcp_2} (a)), (\ref{ldp_2} (a)) and (\ref{hdp_2} (a)), we get the boundaries between different phases.\cite{lakatos03}
The main characteristics of the three dynamical phases, namely the low density 
(LD), high density (HD) and maximal current (MC) phases, that a TASEP 
for single species of hard rod exhibits on the $\alpha-\beta$ plane 
are summarized in Table (\ref{comparison}) \cite{lakatos03,schad10,shaw03,shaw04b,brackley12}.

\begin{widetext}
\begin{center}
\begin{table}
\begin{center}
\begin{tabular}{| c ||c |c |c |  }
  \hline
  \\
  &\textbf{~~~Low Density (LD)} ~~~& ~~~\textbf{High Density (HD)}~~~ &~~~ \textbf{Maximal Current (MC)}~~~ \\
 \\
 \hline
 \\
  Phase boundary condition &$ \alpha<\beta, ~\alpha<\dfrac{p}{1+\sqrt{\ell}} $& $ \beta<\alpha, ~\beta<\dfrac{p}{1+\sqrt{\ell}} $&$ \alpha >\dfrac{p}{1+\sqrt{\ell}},~\beta>\dfrac{p}{1+\sqrt{\ell}} $\\
 \\
 Flux ($ J $)& $ \dfrac{\alpha(p-\alpha)}{p+\alpha(\ell-1)} $&$ \dfrac{\beta(p-\beta)}{p+\beta(\ell-1)} $ &$\dfrac{p}{(\sqrt{\ell}+1)^2}$\\
  \\
 Coverage density ($ \rho_c $) &$ \dfrac{\alpha\ell}{p+\alpha(\ell-1)} $& $\dfrac{p-\beta}{p}$& $\dfrac{\sqrt{\ell}}{(\sqrt{\ell}+1)}$\\
 \\
  \hline
  \end{tabular}
  \caption{Comparison of phase boundary conditions, fluxes and coverage densities in three phases, namely the low density (LD), high density (HD) and maximal current (MC) phases.}
  \label{comparison}
  \end{center}
  \end{table} 
  \end{center}
  \end{widetext}

\subsection{Brief introduction to Internal Ribosome Entry Site (IRES)}

The primary structure of a single DNA strand is a linear sequence of its monomeric 
subunits called nucleotides. The four species of nucleotides that can occur in this 
linear sequence are the analogs of four letters of a language in which nature encodes 
the genetic message. The same message is merely transcribed into the sequence 
of nucleotides on the messenger RNA (mRNA) molecule that is synthesized by a 
RNA polymerase (RNAP) machine using the DNA as a template. Triplets of nucleotides on 
a mRNA are called codons. Following the genetic code, each codon is translated into the corresponding amino 
acid, the monomeric subunits of a protein. Thus, the sequence of codons on a mRNA 
strand serves as the template for the synthesis of the corresponding protein thereby 
translating the genetic message from a language based on 4-letter alphabet to 
another language based on a 20-letter alphabet. The macromolecular machine that 
translates the genetic message is called a {\it ribosome}. From the perspective of 
statistical physics, a ribosome is a molecular motor \cite{chowdhury13a,kolomeisky15} 
that uses the mRNA template also as a track for its step-by-step forward movement. 
After translating each codon a ribosome steps forward to the next codon on the template 
mRNA; thus the step size of a ribosome motor on its track is identical to the linear size 
of a codon. 

Often many ribosomes move simultaneously on the same mRNA track while each 
synthesizes a distinct copy of the same protein whose primary structure is 
encoded in the codon sequence on the template mRNA. Understanding the 
spatio-temporal organization of ribosomes in such traffic-like collective phenomenon  
in terms of a minimal mathematical model \cite{chowdhury05,chou11} was the original 
motivation for formulation of TASEP \cite{macdonald68,macdonald69}. 
Since each ribosome simultaneously covers 
about 30 nucleotides (i.e., 10 codons) each was represented by a rod of length ${\ell} > 1$, 
instead of a particle of length ${\ell}=1$, in the TASEP. The entry and exit of a rod in the 
TASEP capture the initiation and termination, respectively, of the process of translation by 
a ribosome. The hopping of a rod from one lattice site to the next mimics translation of 
successive codons on the mRNA template. Therefore, the average speed of a rod in the 
TASEP corresponds to the rate of elongation of a  protein while the flux of the rods gives 
the overall rate of production of proteins.

Unlike DNA, which forms its iconic double-stranded helix, RNA can adopt wide varieties of 
secondary structures that are believed to have functionally important roles \cite{brierley07,wachter14}. 
Several of these secondary structures are, in fact, ``signals'' for alternative readout of 
the genetic message encoded in its own sequence. Such dynamic alteration of decoding, 
without any alteration of the genetic code, is called ``recoding'' \cite{atkins10,farabaugh97}. 
Recoding phenomena are often referred to as ``programmed'' errors to distinguish these 
from random errors of decoding of genetic message. Except one special type of recoding 
that arises from programmed  error at the level of transcription, almost all types of recoding 
take place via non-canonical translation and appear as programmed translational errors.  

In canonical translation initiation takes place, effectively, at one end of the mRNA template. 
However, one of the non-canonical modes of translation \cite{firth12} can get initiated 
from an ``internal site'' far downstream from the site of canonical initiation (see Fig.\ref{ires_cartoon}). 
For obvious reason, this mode of non-canonical translation is named after the special site of initiation, 
namely, Internal Ribosome Entry Site (IRES) \cite{terenin16}.  The non-canonical routes of 
translation are widely exploited by viruses to hijack the ribosomal machineries of their host 
cells and to evade the anti-viral responses of the host \cite{walsh11,au14}. In response to 
stress, cells are also believed to regulate translation via IRES route \cite{holcik05}. 
Therefore, a deep understanding of the interplay of canonical and non-canonical translation 
of a mRNA can help in developing strategies to combat viral infection. 

The structural features of IRES and the circumstances which promote this mode of non-canonical 
initiation have been the main focus of investigation in the biology literature. One of the common features of IRES, 
observed across all known examples, is that IRES is located on, or just a few nucleotides 
downstrean from, a secondary structure formed by the mRNA template. However, to our 
knowledge, no attention has been paid so far on the consequences of IRES on ribosome traffic.

The main motivation for the two-species exclusion model developed in this paper is to 
explore possible effects of the interference of the flow of two species of ribosomes, 
engaged in canonical and non-canonical translation, on the overall nature of the ribosome 
traffic on a single mRNA. The phenomenon of non-canonical translation via IRES route has 
some similarities with transcriptional interference (TI) where two-species of RNAP motors, 
with their distinct sites of transcription initiation, interfere. Very recently a TASEP-based 
model of TI has been reported elsewhere \cite{ghosh16}. However, as we shall demonstrate 
in this paper, the secondary structure of the mRNA template immediately upstream from 
the IRES, serves as an additional regulator of the extent of interference of the flow of the 
two species of ribosomes. Consequently, much richer varieties of spatio-temporal organization 
of the ribosomes, under different levels of stability of the mRNA secondary structure, are expected 
in the model developed here than those reported in the context of TI in ref.\cite{ghosh16}.

\subsection{Two-species TASEP with hard rods inspired by IRES}

The schematic diagram of the model is shown in Fig. (\ref{ires_model}). This model consists of a one dimensional lattice of size $ L+\ell-1 $ and multiple identical rods of two different species. Size of a rod is $ \ell $ in the units of lattice sites (i.e. it covers $ \ell $ sites simultaneously) and it is assumed to be identical for both species 1 and 2. Here, we denote the position of a rod (either rod of species 1 or 2) by the lattice site at which the left most edge of the rod is located. If the left most edge of the rod is at site $ i $, then, site $ i $ is known as `occupied' site and all remaining sites $ i+1,i+2, \dots, i+\ell-1 $ are `covered' sites. In this model, the whole lattice is divided into three segments: segment \RN{1} consists of site $ i=1 $ to site $ i=L_{1}-1 $ (i.e. $ 1\leq i \leq L_{1}-1 $), segment \RN{2} from site $ i=L_{1} $ to site $ i=L_{2} $ (i.e. $ L_{1}\leq i \leq L_{2} $) and segment \RN{3} from site $ i=L_{2}+1 $ to site $ i=L+\ell-1 $ (i.e. $ L_{2}+1 \leq i \leq L+\ell-1 $).\\

A new rod can enter via two possible pathways: \textit{a}) 
through \textit{canonical} initiation and \textit{b}) through 
\textit{non-canonical} initiation. In the ``canonical initiation'' 
pathway a new rod can attach to the first lattice site (i.e. at $i=1$) 
with rate $ \alpha_1 $, if all the $\ell$ sites $1,2,...{\ell}$ are 
simultaneously empty (i.e. neither occupied nor covered). In the 
alternative ``non-canonical initiation'' pathway a new rod can attach 
to a special internal entry site (i.e. at $ i=i_{s}=L_2+n\ell+m $, 
where $ n,m $ are two integers), with rate $\alpha_{IE}$, if all sites 
from $ i_s $ to $ i_{s}+\ell-1 $ are neither occupied nor covered 
simultaneously, and it will be referred to as ``non-canonical initiation''.
Based on the entry pathway used, the rods can be divided into two 
groups, a) those which enter through the site $ i=1 $ will be referred 
to as of type $rod_1$, and b) those which enter through the site $i=i_s$ 
will be referred to as of type $rod_2$.

In contrast to distinct sites of entry for the two species $rod_1$ and 
$rod_2$, there is only a single downstream site on the lattice from where both 
types of rods exit. If a rod is occupying site $ i=L $ (and is covering 
the sites $i=L+1,L+2,\dots,L+\ell-1$), it can detach from the lattice 
with rate $\beta$, irrespective of whether it is $rod_1$ or $rod_2$. 

Except for initiation and termination, throughout the lattice, a rod can 
jump only in the forward direction, with step size of unity measured in 
the units of lattice spacing. The rate of forward jump is same at all 
sites in the segments \RN{1} and \RN{3}; this rate is $W$ irrespective 
of whether the rod is $rod_1$ or $rod_2$. Jump rate $W_{s}$ inside segment 
\RN{2} is less than $W$ and is assumed to depends on the stability of the 
pseudoknot according to the following expression
\begin{equation}
W_{s}/W=\gamma=\exp(-a~\widehat{\Delta G}),
\label{slow_rate}
\end{equation}
where, $\widehat{\Delta G}$ is proportional to the free energy barrier 
to be overcome by a rod to proceed forward by one step and $a$ is a 
constant of proportionality that is intended to capture the complexities 
of the pseudoknot structure. For all numerical calculations reported in 
this paper the numerical values $ W=1$ and $a=3$ has been chosen.
Thus, the kinetics of the system is governed by the following prescription:\\
{\it Inside segment \RN{1}},\\
\textbf{a:} A new rod ($ rod_1 $) can attach at site $ i=1 $ with rate $ \alpha_1 $, if all sites from $ i=1 $ to $ i=\ell $ are neither covered nor occupied, simultaneously.\\
\textbf{b:} At any site, a rod ($ rod_1 $) can move forward with step size $ 1 $ and rate $ W $, if it is occupying site $ i $ and the target site $ i+\ell $ is not occupied.\\
\textbf{c:} A rod ($rod_{1}$) can exit from segment \RN{1} with rate $ W $, if it is at site $ i=L_1-1 $ and the target site $ i+\ell $ in segment \RN{2} is not occupied. The exclusion process on the segment I can be regarded as a single TASEP for $rod_{1}$ with the effective exit rate $ \beta_{eff1} $ for $rod_1$ from segment \RN{1}.\\
{\it Inside segment \RN{2}},\\
\textbf{a:} The right edge of a rod ($rod_{1}$) can enter in segment \RN{2} with rate $ W $, if its left edge is at site
 $i=L_1-\ell $ 
and the target site $ i+\ell $ is not occupied. The exclusion process in segment II can be regarded as a single TASEP for $rod_{1}$ with the effective rate $ \alpha_{eff2} $ for the entry of $ rod_1 $ into the segment \RN{2}.\\
\textbf{b:} At any site a rod ($ rod_1 $) can move forward with step size $ 1 $ and rate $ W_{s} $, if it is occupying site $ i $ and the target site $ i+\ell $ is  not occupied.\\
\textbf{c:} A rod ($rod_{1}$) can exit from segment \RN{2} with rate $ W_s $, if its left edge is at site $i=L_2 $ and the target site $ i+\ell $ in segment \RN{3} is not occupied by any other rod. Thus the exclusion process in segment II can be regarded as a TASEP with the effective rate $ \beta_{eff2} $ of exit of $rod_1$ from segment \RN{2}.\\
{\it Inside segment \RN{3}},\\
\textbf{a:} A $rod_{1}$ can enter segment \RN{3} from segment II with rate $ W_s $, if its left edge is at site $i=L_2-\ell+1 $ and the target site $ i+\ell $ is not occupied by any other rod. Thus, the effective rate of entry of $rod_1$ into segment \RN{3} is $ \alpha_{NIE} $.\\
\textbf{b:} Inside this segment there is one special site at $ i=i_s $ to which a new $ rod_2 $ can attach with rate $\alpha_{IE}$ if none of the sites from $ i_s $ to $ i_{s}+\ell-1 $ are covered. Thus, the exclusion process in segment III can be regarded as a TASEP of mixed population of $rod_{1}$ and $rod_{2}$ with the total entry rate $\alpha_{eff3}$.\\
\textbf{c:} A rod ($ rod_1 $ or $ rod_2 $), can completely detach from the lattice with rate  $ \beta $, if it is occupying site $ i=L $ and covering remaining sites $ i=L+1,\dots, L+\ell-1 $. The rate of detachment is assumed to be identical for both $ rod_1 $ and $ rod_2 $.\\
\textbf{d:} At all other sites, a rod ($ rod_1 $ or $ rod_2 $) can move forward with step size $ 1 $ and rate $ W $, if its left edge is located at site $ i $ and the target site $ i+\ell $ is not occupied by any other rod.\\
The effective rate constants and their defining expressions are given in table 
\ref{rate_1}. Note that the exclusion process in each of the segments I and II are single-species exclusion process whereas that in segment III is a two-species exclusion process.  However, for simplicity, we assume the same hopping rate $W$ and exit rate $\beta$ in this segment for both species of the rods.

\begin{table}
\begin{center}
\begin{tabular}{| c | c | }
  \hline
  Rate constant & Expression  \\
  \hline
  \hline
  $\alpha_{eff2} $& $\alpha_{eff2}=W \rho_{c1}/\ell$ \\
  \hline
  $\alpha_{NIE} $& $\alpha_{NIE}=W_s \rho_{c2}/\ell$ \\
  \hline
  $\alpha_{eff3} $& $\alpha_{eff3}=\alpha_{NIE}+\alpha_{IE}$ \\
  \hline
  $\beta_{eff1} $ & $ \beta_{eff1}=WP_1(\underline{L_1-1} \vert L_1+\ell-1) $ \\
  \hline
  $\beta_{eff2} $ & $ \beta_{eff2}=W_sP_1(\underline{L_2} \vert L_2+\ell) $ \\
  \hline
  \end{tabular}
  \caption{The effective rate constants and their defining expressions.}
  \label{rate_1}
  \end{center}
  \end{table} 
\subsection{Two models of Non-canonical initiation}
\noindent$\bullet${\it Signal Independent Initiation}:\\
In the first model, one assumes that no correlation exists between the canonical and non-canonical initiation so that the rates $\alpha_1$ and $\alpha_{IE}$ 
of the entry of $rod_{1}$ and $rod_{2}$ in the segments I and III, 
respectively are two independent parameters. \\
\noindent$\bullet${\it Signal Dependent Initiation}:\\
In the alternative model, the rates $\alpha_1$ and $\alpha_{IE}$ of canonical 
and non-canonical initiation, respectively, cannot be treated as independent 
parameters. Instead, the rates of entry of $rod_{1}$ and $rod_{2}$ are 
controlled by an external signal according to the prescription 
\begin{equation}
\alpha_1=s~\alpha,
\label{alpha_1}
\end{equation}
and\\
\begin{equation}
\alpha_{IE}=(1-s)~\alpha.
\label{alpha_2}
\end{equation}
where, $\alpha$ is the total rate of initiation and an external signal 
decides the fraction, denoted by $`s`$, of initiation events that occur 
via the canonical route. We compare the results for these two models in the concluding section.

\subsection{Quantities of interest}

Let $ P_{\mu}(i,t)$ denote the probability of finding a rod  of type $ \mu (\mu=1,2) $ at site $ i $ at time $ t $.  We use the symbols $J_{seg1}, J_{seg2}$ and $J_{seg3}$ to denote the 
net flux of the rods in the segments \RN{1}, \RN{2} and \RN{3}, respectively. 
Under steady state conditions, $J_{seg\mu}$ ($\mu=1,2,3$) would be given 
by either (\ref{ldp_2})(a) or (\ref{hdp_2})(a) or (\ref{mcp_2})(a) depending 
on whether the segment under consideration is in the LD, or HD or MC phase; 
the entry and exit rates to be used in these equations are the effective rates 
for the respective segments, namely $\alpha_{1}, \beta_{eff1}$ for  \RN{1}, 
$\alpha_{eff2},\beta_{eff2}$ for  \RN{2} and $\alpha_{eff3}, \beta$ for  \RN{3}. Under steady state condition, flux in the three segments must satisfy the 
following conditions:
\begin{equation}
J_{seg1}=J_{seg2},
\label{flux_11}
\end{equation}
and
\begin{equation}
J_{seg3}=J_{seg1}+\alpha_{IE}P(\underbrace{0, \dots,0}_{\ell}).
\label{flux_22}
\end{equation}
where $ P(\underbrace{0, \dots,0}_{\ell}) $ is the probability that none  
of the sites from $i=i_s $ to $ i=i_s+\ell-1 $ are covered.

The net flux gets contributions from the forward movemetns of both types 
of rods. The fluxes $J_{1}$ and $J_{2}$ of $rod_{1}$ and $rod_{2}$, 
respectively, which are the number of RNA molecules synthesized per unit 
time in the canonical and non-canonical translation processes, are 
defined by 
\begin{equation}
J_{\mu}=P_{\mu}(L,t)\beta ~~~~~~~  (\mu=1,2).
\label{flux}
\end{equation}

Since, the system consists of three segments where each has some distinct 
kinetics of the rods, the coverage densities in these three segments can 
be different from each other. We use the symbols $ \rho_{c1}, \rho_{c2}$ 
and $ \rho_{c3} $ for the coverage densities in segments \RN{1}, \RN{2} and 
\RN{3}, respectively. The overall spatio-temporal organization of the rods, arising from the inhomogeneous density profiles, but uniform flux, in the three-segment system,  can be studied by displaying  all possible composite phases on the phase diagram. Each composite phase comprises of the three phases that can exist in three segments. In principle, each segment can exist in one of the three possible phases i.e. low density (LD), high density (HD) and maximal current (MC) phase.  The composite phase diagrams will be denoted by the symbols $X/Y/Z$, where $X,Y, Z$ correspond to the phases in the segments I, II, II, respectively, and where each of the three symbols $X,Y,Z$ can be either LD or HD or MC.


\subsection{Master equations}
We define $ p(\underline{i} \mid i+\ell) $ as the conditional probability of finding site $ i+\ell $ empty, if site $ i $ is given to be occupied. It is straightforward to show that \cite{basu07,gccr09}
\begin{equation}
p(\underline{i} \mid i+\ell)=\dfrac{1-\rho \ell}{1+\rho-\rho \ell},
\label{conditional_1}
\end{equation}
where, $ \rho $ is the occupational density (i.e., number density) of the rods in the system.
Similarly, we define $P(\underbrace{0, \dots,0}_{\ell})$ as the probability that all sites from $ i=i_s $ to $ i=i_s+\ell-1 $ are neither occupied nor covered, simultaneously. Expression of  $P(\underbrace{0, \dots,0}_{\ell})$ in terms of occupational density $ \rho_3 $ of \RN{3} (when $ i_s=L_2+n\ell+m $ and $ n \geq 1 $), is given by (see the appendix for full derivation of (Eq. \ref{cond_prob_1}, \ref{cond_prob_2} and  \ref{cond_prob_3})),\\
\begin{equation}
P(\underbrace{0, \dots,0}_{\ell})= \dfrac{2(1-\rho_{3})}{2(1-\rho_{3})(1+\rho_{3}^\prime) + \rho_{3}\ell(2+\rho_{3}^\prime)}.
\label{conditional_2}
\end{equation}
where, $\rho_{3}^\prime=\rho_{3}(\ell-1)$.\\

Master equations, under mean field approximation (MFA), corresponding to the stochastic kinetics explained above are given by the following expressions:\\
(A) For $ rod_1 $\\
{\it in segment \RN{1},}\\
\textbf{a:} At site $ i=1 $,
\begin{equation}
\dfrac{dP_{1}(i,t)}{dt}=\alpha_1\Big[1-\sum_{k=1}^{\ell}P_{1}(k,t)\Big]-P_{1}(i,t)p(\underline{i} \mid i+\ell)W, 
\label{site1rod1}
\end{equation}
\textbf{b:} At all remaining sites,
\begin{equation}
\begin{split}
\dfrac{dP_{1}(i,t)}{dt} & =P_{1}(i-1,t)p(\underline{i-1} \mid i+\ell-1)W \\
& -P_{1}(i,t)p(\underline{i} \mid i+\ell)W, 
\end{split}
\label{seg1rod1}
\end{equation}
{\it In segment \RN{2},}\\
\begin{equation}
\begin{split}
\dfrac{dP_{1}(i,t)}{dt} & =P_{1}(i-1,t)p(\underline{i-1} \mid i+\ell-1)W_s \\
& -P_{1}(i,t)p(\underline{i} \mid i+\ell)W_s,
\end{split}
\label{seg2rod1}
\end{equation}
{\it In segment \RN{3},}\\
\textbf{a:} At site $ i=L $,
\begin{equation}
\dfrac{dP_{1}(i,t)}{dt} =P_{1}(i-1,t)W -P_{1}(i,t)\beta,
\label{siteNrod1}
\end{equation}
\textbf{b:} At all remaining sites,
\begin{equation}
\begin{split}
\dfrac{dP_{1}(i,t)}{dt} & =P_{1}(i-1,t)p(\underline{i-1} \mid i+\ell-1)W \\
& -P_{1}(i,t)p(\underline{i} \mid i+\ell)W.
\end{split}
\label{seg3rod1}
\end{equation}
(B) For $ rod_2 $\\
{\it if $ i < i_s $,}\\
\begin{equation}
P_{2}(i,t)=0,
\label{rod2seg1,2}
\end{equation}
{\it in segment \RN{3}}\\

\textbf{a:} At site $ i=i_{s}$,
\begin{equation}
\begin{split}
\dfrac{dP_{2}(i,t)}{dt}&=\alpha_{IE}P(\underbrace{0, \dots,0}_{\ell})\\
& -P_{2}(i,t)p(\underline{i} \mid i+\ell)W, 
\end{split}
\label{site_isrod2}
\end{equation}

\textbf{b:} At site $ i=L $,
\begin{equation}
\dfrac{dP_{2}(i,t)}{dt}=P_{2}(i-1,t)W -P_{2}(i,t)\beta,
\label{siteLrod2}
\end{equation}
\textbf{c:} At all remaining sites ,
\begin{equation}
\begin{split}
\dfrac{dP_{2}(i,t)}{dt} & =P_{2}(i-1,t)p(\underline{i-1} \mid i+\ell-1)W \\
& -P_{2}(i,t)p(\underline{i} \mid i+\ell)W. 
\end{split}
\label{seg3rod2}
\end{equation}


\section{Results}
We have followed two different approaches for  studying the combined effect of initiation rates $ \alpha_1, \alpha_{IE}$ and the jump rate $W_{s}$ in segment \RN{2} (or, more precisely, the ratio $\gamma =W_{s}/W$) on the spatio-temporal organization of the two species of rods in the system. Our first approach is based on an analysis of the master equations  formulated above under MFA. This analysis is primarily analytical derivation of the phase boundaries and fluxes in the steady states of the system.

In our second approach, we carry out extensive Monte-Carlo (MC) simulations of the same model. In the MC simulations, starting from an arbitrary initial state, the state of the system was updated for sufficiently large number of MC steps to achieve steady state after which the relevant data were collected over the  next $ 1 \times 10^{5} $ MC steps. In order to convert the rates (with units s$^{-1}$) of the various processes into the corresponding dimensionless probabilities we have  used the prescription
\begin{equation}
p_{k}=k~dt
\label{rate_conversion}
\end{equation}
where, $p_{k}$ is the probability corresponding to an arbitrary rate constant $ k $ and $ dt $ is a small time step; for all the numerical data presented graphically in this paper  we have chosen $dt = 0.01$s. 

\subsection{Analytical derivations of phase boundaries and fluxes}

We can model our system as a combination of three TASEP, where a TASEP on the left (right) of a given segment acts, effectively, as a reservoir from where rods enter (to which rods exit) the given segment. Therefore, at first sight, it may appear that the entire system could be found in $ 3^{3} =27$ possible phases. But, because of some symmetry requirements and steady state conditions not all $ 27 $ phases can exist in the system.

The steady-state flux inside segments \RN{1} and \RN{2} should always be identical. 
If MC phase exist in segment \RN{1}, the flux inside this segment should be equal to, $ J_{seg1}=1/(\sqrt{\ell}+1)^2 $. However, if segment \RN{2} is in MC phase, $ J_{seg2}= W_s/(\sqrt{\ell}+1)^2<1/(\sqrt{\ell}+1)^2$, when $ W_{s}<1 $.
Therefore, those composite phases are ruled out, where the segment \RN{1} is in MC phase while the segment \RN{2} is not. 
Similarly, if both the segments \RN{1} and \RN{2} are in HD phase, the probability of finding segment \RN{3} in LD phase is approaching towards zero value, when $ \alpha_{IE} \neq 0 $.
Taking into account all the conditions arising from symmetry considerations, the model developed in this paper can exist in only seven possible phase. We characterize here only those seven phases based on analytical treatment as well as MC simulations.\\

We now begin presentation of our results by analytically deriving the phase boundaries of the system.  

\textbf{Phase 1: LD/LD/LD phase}

Since all the segments are in LD phases, this composite phase of the system is specified by 
the following conditions,\\
\begin{eqnarray}
\alpha_1  & < \frac{1}{\sqrt{{\ell}}+1} , ~~~~\alpha_1 < \beta_{eff1}, \nonumber \\
\alpha_{eff2} & < \frac{W_s}{\sqrt{{\ell}}+1} , ~~~~\alpha_{eff2} < \beta_{eff2} \nonumber, \\
\alpha_{eff3} & < \frac{1}{\sqrt{{\ell}}+1} , ~~~~\alpha_{eff3} < \beta.
\label{ldldld_1}
\end{eqnarray}
Under steady state condition the fluxes of the rods that follow from the equations (\ref{ldp_2})(a), are given by\\
\begin{eqnarray}
J_{seg1} &=& \frac{\alpha_1(1-\alpha_1)}{1+\alpha_1(\ell-1)} , ~~(a)\nonumber \\
 J_{seg2} &=& \frac{\alpha_{eff2}(W_s-\alpha_{eff2})}{W_s+\alpha_{eff2}(\ell-1)} , ~~(b) \nonumber \\
 J_{seg3} &=& \frac{\alpha_{eff3}(1-\alpha_{eff3})}{1+\alpha_{eff3}(\ell-1)} , ~~(c)\nonumber\\
 \label{ldldld_2J}
\end{eqnarray}
while, from  the equations (\ref{ldp_2})(b), the corresponding coverage densities are given by 
\begin{eqnarray}
\rho_{c1}&=&\frac{\alpha_1\ell}{1+\alpha_1(\ell-1)}, ~~(a)\nonumber \\
\rho_{c2}&=&\frac{\alpha_{eff2} \ell}{W_s+\alpha_{eff2}(\ell-1)}, ~~(b) \nonumber \\
\rho_{c3}&=&\frac{\alpha_{eff3} \ell}{1+\alpha_{eff3}(\ell-1)}.~~(c)\nonumber\\.
 \label{ldldld_2rho}
\end{eqnarray}

Now imposing the condition (\ref{flux_11}), i.e., equating (\ref{ldldld_2J})(a) and (\ref{ldldld_2J})(b), we get\\
\begin{eqnarray}
\frac{\alpha_{1}(1-\alpha_{1})}{1+\alpha_{1}(\ell-1)}-\frac{\alpha_{eff2}(W_s-\alpha_{eff2})}{W_s+\alpha_{eff2}(\ell-1)}=0,
\label{ldldld_3}
\end{eqnarray}
solving which for the unknown $\alpha_{eff2}$ we get,\\

\begin{eqnarray}
\alpha_{eff2}=\frac{-t_1-\sqrt{t_1^2-4t_2}}{2 t_3},
\label{ldldld_4}
\end{eqnarray}
where,
\begin{eqnarray}
t_1=-(\alpha_1^2 \ell-\alpha_1^2+\alpha_1 \ell W_s-\alpha_1 \ell-\alpha_1 W_s+\alpha_1 +W_s), \nonumber \\
\label{t1}
\end{eqnarray}
\begin{eqnarray}
t_2=(\alpha_1 \ell -\alpha_1 +1) (\alpha_1 W_s-\alpha_1^2 W_s),
\label{t2}
\end{eqnarray}
\begin{eqnarray}
t_3=\alpha_1 \ell -\alpha_1 +1.
\label{t3}
\end{eqnarray}
Note that of the two solutions of the quadratic equation we have chosen the one with the negative sign to be the physical solution because $\alpha_{eff2}$ should be less than $ W_s/(\sqrt{{\ell}}+1) $.\\
 
Combining Eqn. (\ref{ldldld_4}) and the condition (\ref{ldldld_1}), 
we get the maximum physically admissible value of the $\alpha_{1}$ 
upto which both the segments \RN{1} and \RN{2} would exist in LD phases. 
Thus, substituting Eq. (\ref{ldldld_4}) into (\ref{ldldld_1}) we get
\begin{eqnarray}
\frac{-t_1-\sqrt{t_1^2-4t_2}}{2 t_3} < \frac{W_s}{(\sqrt{\ell}+1)},
\label{ldldld_comb}
\end{eqnarray}
which, for example, for $\ell=10$ reduces to the condition
\begin{eqnarray}
\alpha_1 < \frac{t_{11}-t_{12}}{18},
\label{ldldld_a1_cond}
\end{eqnarray}
for the occurrence of LD phase in both the segments \RN{1} and \RN{2} where
\begin{eqnarray}
t_{11}=2 \sqrt{10} W_s-11 W_s+9,
\label{t11}
\end{eqnarray}
\begin{eqnarray}
t_{12}=\sqrt{(2 \sqrt{10} W_s-11 W_s+9)^2-(44 - 8 \sqrt{10}) W_s}. \nonumber \\
\label{t12}
\end{eqnarray}
Substituting the expression (\ref{ldldld_4}) for $\alpha_{eff2}$ into those of $J_{seg2}$ in (\ref{ldldld_2J}) and $\rho_{c2}$ in (\ref{ldldld_2rho}) we get the flux $J_{seg2}$ and coverage density $\rho_{c2}$ in terms of $\alpha_{1}$ and $W_{s}$,\\
\begin{eqnarray}
\rho_{c2}=\frac{\ell (-t_1-\sqrt{t_1^2-4t_2})}{ 2t_3 \bigg(W_s+\frac{(\ell-1) (-t_1-\sqrt{t_1^2-4t_2})}{2t_3}\bigg)}.
\label{ldldld_5}
\end{eqnarray}
Furthermore, substituting the expressions thus obtained for $\alpha_{eff2}$ and $\rho_{c2}$ into
\begin{eqnarray}
\alpha_{NIE}=W_s\frac{\rho_{c2}}{\ell},
\end{eqnarray}
we get the expression 
\begin{eqnarray}
\alpha_{NIE}=\frac{W_s (-t_1-\sqrt{t_1^2-4t_2})}{ 2t_3 \bigg(W_s+\frac{(\ell-1) (-t_1-\sqrt{t_1^2-4t_2})}{2t_3}\bigg)},
\label{ldldld_6}
\end{eqnarray}
for $\alpha_{NIE}$, and hence 
\begin{eqnarray}
\alpha_{eff3} &=& \alpha_{IE} + \alpha_{NIE} \nonumber \\
&=& \alpha_{IE} + \frac{W_s (-t_1-\sqrt{t_1^2-4t_2})}{ 2t_3 \bigg(W_s+\frac{(\ell-1) (-t_1-\sqrt{t_1^2-4t_2})}{2t_3}\bigg)}.
\label{ldldld_7}
\end{eqnarray}
From condition (\ref{ldldld_1}), 
\begin{eqnarray}
\beta > \alpha_{eff3}&=&\alpha_{IE} \nonumber \\ 
&+& \frac{W_s (-t_1-\sqrt{t_1^2-4t_2})}{ 2t_3 \bigg(W_s+\frac{(\ell-1) (-t_1-\sqrt{t_1^2-4t_2})}{2t_3}\bigg)}.
\label{ldldld_9}
\end{eqnarray}
The canonical and non-canonical fluxes in the LD/LD/LD phase are given by the expressions
\begin{eqnarray}
 J_1 &=& J_{seg1} =J_{seg2} =\dfrac{\alpha_{eff2}(W_s-\alpha_{eff2})}{(W_s+\alpha_{eff2}(\ell-1))},  \nonumber  \\
 {\rm and}\\
 J_2 &=&\dfrac{\alpha_{eff3}(1-\alpha_{eff3})}{(1+\alpha_{eff3}(\ell-1))}- \dfrac{\alpha_{eff2}(W_s-\alpha_{eff2})}{(W_s+\alpha_{eff2}(\ell-1))}. 
 \label{ldldld_61}
 \end{eqnarray}

\textbf{Phase 2: LD/LD/HD phase}
The LD/LD/HD phase is specified by the following conditions, 
\begin{eqnarray}
\alpha_1  &<& \frac{1}{\sqrt{{\ell}}+1} , ~~~~\alpha_1 < \beta_{eff1}, \nonumber \\
\alpha_{eff2} &<& \frac{W_s}{\sqrt{{\ell}}+1} , ~~~~\alpha_{eff2} < \beta_{eff2}, \nonumber \\
\beta &<& \frac{1}{\sqrt{{\ell}}+1} , ~~~~\beta < \alpha_{eff3},
\label{ldldhd_1}
\end{eqnarray}
in terms of the effective rates;  
the corresponding steady-state fluxes that follow from the equations (\ref{ldp_2})(a) and (\ref{hdp_2})(a), would be
\begin{eqnarray}
J_{seg1} &=& \frac{\alpha_1(1-\alpha_1)}{1+\alpha_1(\ell-1)} , ~~(a)\nonumber \\
 J_{seg2} &=& \frac{\alpha_{eff2}(W_s-\alpha_{eff2})}{W_s+\alpha_{eff2}(\ell-1)} , ~~(b) \nonumber \\
  J_{seg3} &= &\frac{\beta(1-\beta)}{1+\beta(\ell-1)} , ~~(c)\nonumber\\
 \label{ldldhd_2J}
\end{eqnarray}
while, from  the equations (\ref{ldp_2})(b) and (\ref{hdp_2})(b) the corresponding coverage densities would be given by 
\begin{eqnarray}
\rho_{c1}&=&\frac{\alpha_1\ell}{1+\alpha_1(\ell-1)}, ~~(a)\nonumber \\
\rho_{c2}&=&\frac{\alpha_{eff2} \ell}{W_s+\alpha_{eff2}(\ell-1)}, ~~(b) \nonumber \\
\rho_{c3}&=&1-\beta.~~~~~~~~(c)\nonumber\\
 \label{ldldhd_2rho}
\end{eqnarray}
Now imposing the condition (\ref{flux_11}), i.e., equating (\ref{ldldhd_2J})(a) and (\ref{ldldhd_2J})(b), and
solving for the unknown $\alpha_{eff2}$ we get the same expression (\ref{ldldld_4}). 
This, however, is not surprising because $\alpha_{eff2}$ for both the phases 1 and 2 is a characteristics of the 
LD/LD interface.

Next imposing Eq. (\ref{flux_22}), i.e., equating the fluxes of segment \RN{2} and \RN{3} we get,\\
\begin{equation}
\frac{\beta(1-\beta)}{1+\beta(\ell-1)}=\frac{\alpha_{eff2}(W_s-\alpha_{eff2})}{W_s+\alpha_{eff2}(\ell-1)} + \alpha_{IE}P(\underbrace{0, \dots,0}_{\ell}).
\label{ldldhd_4}
\end{equation}
Substituting the expression 
\begin{equation}
P(\underbrace{0, \dots,0}_{\ell})=\dfrac{2\beta}{2\beta(1+\beta^\prime) +(1-\beta)\ell(2+\beta^\prime)}.
\label{ldldhd_cond}
\end{equation}
(see the derivation of Eq. (\ref{hd_1}) in the Appendix for the LD/LD/HD phase), where, $\beta^\prime = (1-\beta)(\ell-1)$,
into Eq.(\ref{ldldhd_4}) we get
\begin{eqnarray}
\frac{\beta(1-\beta)}{1+\beta(\ell-1)} =\frac{\alpha_{eff2}(W_s-\alpha_{eff2})}{W_s+\alpha_{eff2}(\ell-1)}  \nonumber \\
 + \dfrac{2\alpha_{IE}\beta}{2\beta(1+\beta^\prime) +(1-\beta)\ell(2+\beta^\prime)}.
\label{ldldhd_5}
\end{eqnarray}

Solving this quartic equation for $\beta$, we get four solutions of which 
two are negative solutions and, hence, inadmissible.  Out of two positive 
solutions we have chosen the one with the negative sign to be the physical 
solution because, as dictated by (\ref{ldldhd_1}), $ \beta $ must be less 
than $1/({\sqrt{{\ell}}+1})$.
Since the general expression for $\beta$ is too long to be reproduced here, the expression of $ \beta $ for the special values of the parameter set: $ W_s=0.9, \ell=10, \alpha_{IE}= 0.15$ is given in the appendix \ref{app-Phase2}.

Substituting the expression (\ref{ldldld_4}) for $\alpha_{eff2}$ in the solution for $ \beta $ thus obtained, we get the relation between  $ \alpha_{1} $ and $ \beta $ in terms of $ W_s $, $ \ell $ and $ \alpha_{IE} $. This relation gives us a transition line which separates the LD/LD/HD from HD/HD/HD phase. During this transition, density of segment \RN{1} and \RN{2} changes discontinuously. 
Thus, the conditions  (\ref{ldldld_9}) and (\ref{ldldhd_1}), written in terms of the effective rates, now reduce to  
\begin{eqnarray}
\alpha_{IE} &+ &\frac{W_s (-t_1-\sqrt{t_1^2-4t_2})}{ 2t_3 \bigg(W_s+\frac{(\ell-1) (-t_1-\sqrt{t_1^2-4t_2})}{2t_3}\bigg)} > \beta > \Big[ -s_{0}  \nonumber \\
&-&0.5 \sqrt{s_0-s_1-(s_2/s_3)-s_4}  \nonumber \\
&-&0.5 \sqrt{s_5-s_{6}+(s_2/s_3)+s_4-(s_7/s_{8})}\Big]  
\label{ldldhd_7}
\end{eqnarray}
for the composite phase LD/LD/HD of the system where $s_{\mu} (\mu=0,1, \cdots 8)$ are expressed in terms of 
the basic rate constants of the model (see appendix \ref{app-Phase2} for an example).  

Finally, in the LD/LD/HD phase the canonical and non-canonical fluxes are given by \\
 \begin{eqnarray}
 J_1&=&J_{seg1} =J_{seg2} =\dfrac{\alpha_{eff2}(W_s-\alpha_{eff2})}{(W_s+\alpha_{eff2}(\ell-1))}, ~~(a) \nonumber  \\
 {\rm and} \nonumber \\
 J_2&=& \dfrac{\beta(1-\beta)}{(1+\beta(\ell-1))}-\dfrac{\alpha_{eff2}(W_s-\alpha_{eff2})}{(W_s+\alpha_{eff2}(\ell-1))}. ~~(b) \nonumber \\
 \label{ldldhd_8}
 \end{eqnarray}
respectively, where $\alpha_{eff2}$ is given by (\ref{ldldld_4}).


\noindent \textbf{Phase 3: LD/LD/MC phase}\\
The LD/LD/MC phase is specified by the conditions,\\
\begin{eqnarray}
 \alpha_1  &<& \frac{1}{\sqrt{\ell}+1} , ~~~~\alpha_1 < \beta_{eff1}, \nonumber \\
\alpha_{eff2} &<& \frac{W_s}{\sqrt{\ell}+1} , ~~~~\alpha_{eff2} < \beta_{eff2}, \nonumber \\
\beta &>& \frac{1}{\sqrt{\ell}+1} , ~~~~\alpha_{eff3}  > \frac{1}{\sqrt{\ell}+1},
\label{ldldmc_1}
\end{eqnarray}
in terms of the effective rates;  the corresponding steady-state fluxes that follow from the equations (\ref{ldp_2})(a) and (\ref{mcp_2})(a), would be
\begin{eqnarray}
J_{seg1} &=& \frac{\alpha_1(1-\alpha_1)}{1+\alpha_1(\ell-1)} , ~~(a)\nonumber \\
 J_{seg2} &=& \frac{\alpha_{eff2}(W_s-\alpha_{eff2})}{W_s+\alpha_{eff2}(\ell-1)} , ~~(b) \nonumber \\
  J_{seg3} & = &  \frac{1}{(\sqrt{\ell}+1)^{2}}, ~~(c)\nonumber\\
 \label{ldldmc_2J}
\end{eqnarray}
while, from  the equations (\ref{ldp_2})(b) and (\ref{mcp_2})(b) the corresponding coverage densities would be given by 
\begin{eqnarray}
\rho_{c1}&=&\frac{\alpha_1\ell}{1+\alpha_1(\ell-1)}, ~~(a)\nonumber \\
\rho_{c2}&=&\frac{\alpha_{eff2} \ell}{W_s+\alpha_{eff2}(\ell-1)}, ~~(b) \nonumber \\
\rho_{c3}&=& \frac{\sqrt{\ell}}{\sqrt{\ell}+1}.~~~~~~~~(c)\nonumber\\.
 \label{ldldmc_2rho}
\end{eqnarray}
Imposition of the condition (\ref{flux_11}), as before, yields the expression (\ref{ldldld_4}) for $\alpha_{eff2}$.
Now from Eq. (\ref{flux_22}), i.e., equating the fluxes of segment \RN{2} and \RN{3}i, we get
\begin{eqnarray}
 \frac{1}{(\sqrt{\ell}+1)^{2}}=\frac{\alpha_{eff2}(W_s-\alpha_{eff2})}{W_s+\alpha_{eff2}(\ell-1)} + \alpha_{IE}P(\underbrace{0, \dots,0}_{\ell}), \nonumber \\
\label{ldldmc_4}
\end{eqnarray}
substitution of 
\begin{equation}
 P(\underbrace{0, \dots,0}_{\ell})=\dfrac{4}{2(\ell+1) +\ell(\ell+3)},
\label{ldldmc_cond}
\end{equation}
for LD/LD/MC phase (see the derivation of Eq. (\ref{mc_1}) in the Appendix) 
 into Eq.(\ref{ldldmc_4}) leads to
\begin{eqnarray}
 \frac{1}{(\sqrt{\ell}+1)^{2}}  &=& \frac{\alpha_{eff2}(W_s-\alpha_{eff2})}{W_s+\alpha_{eff2}(\ell-1)}  \nonumber \\
 &+& \dfrac{4\alpha_{IE}}{2(\ell+1) +\ell(\ell+3)}.
\label{ldldmc_5}
\end{eqnarray}

Solving this quadratic equation (\ref{ldldmc_4}) we get two solutions 
for of $ \alpha_{eff2}$;  we have chosen the solution with the negative 
sign to be the physical solution because, as the eq.(\ref{ldldmc_1}) 
requires, $\alpha_{eff2}$ must be less than $ W_s/(\sqrt{{\ell}}+1)$. 
The Eq.(\ref{ldldmc_4}) leads to
\begin{eqnarray}
\alpha_{eff2} = \frac{-t_4-\sqrt{t_4^2-4t_5t_6}}{2 t_6},
\label{ldldmc_6}
\end{eqnarray}
The general expresions for $ t_4,t_5 \text{ and } t_6 $ are given in the appendix \ref{app-Phase3} (see Eq. \ref{t4},~\ref{t5} and \ref{t6}).\\
The expression (\ref{ldldmc_6}) of $\alpha_{eff2}$, that we get here 
in terms of $ W_s $, $ \ell $ and $ \alpha_{IE} $, guarantees that 
the segment \RN{2} is in LD phase while \RN{3} is in MC phase.
From Eqn. (\ref{ldldld_a1_cond}), we get the maximum physically 
admissible value of the $ \alpha_{1} $ to get both the segments 
\RN{1} and \RN{2} in LD phase and from condition (\ref{ldldmc_1}),
\begin{equation}
\beta > \frac{1}{\sqrt{\ell}+1}.
\label{ldldmc_7}
\end{equation}
The expressions of canonical and non-canonical flux for LD/LD/MC phase 
are given by
\begin{eqnarray}
 J_1 &=&J_{seg1} =J_{seg2} =\dfrac{\alpha_{eff2}(W_s-\alpha_{eff2})}{(W_s+\alpha_{eff2}(\ell-1))},  \nonumber  \\
{\rm and} \nonumber \\
 J_2&=&\frac{1}{(\sqrt{\ell}+1)^2}-\dfrac{\alpha_{eff2}(W_s-\alpha_{eff2})}{(W_s+\alpha_{eff2}(\ell-1))},
 \label{ldldmc_8}
 \end{eqnarray}
respectively, where $\alpha_{eff2}$ is given by (\ref{ldldld_4}).

\noindent \textbf{Phase 4: HD/HD/HD phase}

In terms of the effective rates, the HD/HD/HD phase is specified by the conditions,
\begin{eqnarray}
\beta_{eff1}  &<& \frac{1}{\sqrt{{\ell}}+1} , ~~~~ \beta_{eff1} < \alpha_1,  \nonumber \\
\beta_{eff2} &<& \frac{W_s}{\sqrt{{\ell}}+1} , ~~~~ \beta_{eff2} < \alpha_{eff2},  \nonumber \\
\beta &<& \frac{1}{\sqrt{{\ell}}+1} , ~~~~ \beta < \alpha_{eff3}.
\label{hdhdhd_1}
\end{eqnarray}
The corresponding fluxes of the rods in the three segments are  
\begin{eqnarray}
J_{seg1} &=& \frac{\beta_{eff1}(1-\beta_{eff1})}{1+\beta_{eff1}(\ell-1)}, ~~(a)\nonumber \\
  J_{seg2} & = &\frac{\beta_{eff2}(W_s-\beta_{eff2})}{W_s+\beta_{eff2}(\ell-1)} , ~~(b) \nonumber \\
  J_{seg3} &= &\frac{\beta(1-\beta)}{1+\beta(\ell-1)}, ~~(c)\nonumber\\
 \label{hdhdhd_2J}
\end{eqnarray}
while, the corresponding coverage densities are 
\begin{eqnarray}
 \rho_{c1}&=&1-\beta_{eff1}, ~~(a)\nonumber \\
 \rho_{c2}&=&1-(\beta_{eff2}/W_s), ~~(b) \nonumber \\
\rho_{c3}&=&1-\beta.~~~(c)\nonumber\\.
 \label{hdhdhd_2rho}
\end{eqnarray}

Imposing the condition (\ref{flux_22}), i.e., equating the fluxes of segment \RN{2} and \RN{3} we get,\\
\begin{eqnarray}
\frac{\beta(1-\beta)}{1+\beta(\ell-1)}=\frac{\beta_{eff2}(W_s-\beta_{eff2})}{W_s+\beta_{eff2}(\ell-1)} + \alpha_{IE}P(\underbrace{0, \dots,0}_{\ell}), \nonumber \\
\label{hdhdhd_2}
\end{eqnarray}
for HD/HD/HD phase (see Eq. (\ref{hd_1}) in the Appendix),\\
\begin{equation}
P(\underbrace{0, \dots,0}_{\ell})=\dfrac{2\beta}{2\beta(1+\beta^\prime) +(1-\beta)\ell(2+\beta^\prime)}.
\label{hdhdhd_cond}
\end{equation}
where, $\beta^\prime = (1-\beta)(\ell-1) $.\\
Substituting Eq.(\ref{hdhdhd_cond}), in Eq.(\ref{hdhdhd_2}) it becomes,\\
\begin{eqnarray}
\frac{\beta(1-\beta)}{1+\beta(\ell-1)}  =\frac{\beta_{eff2}(W_s-\beta_{eff2})}{W_s+\beta_{eff2}(\ell-1)}  \nonumber \\
 + \dfrac{2\alpha_{IE}\beta}{2\beta(1+\beta^\prime) +(1-\beta)\ell(2+\beta^\prime)}.
 \label{hdhdhd_3}
\end{eqnarray}

Solving this quadratic equation (\ref{hdhdhd_3}) we get the expression of $ \beta_{eff2} $, in terms of $ \beta $, $ W_s $ and $ \ell $: \\
\begin{eqnarray}
\beta_{eff2}=\frac{-h_2-\sqrt{h_2^2-4h_1h_3}}{2h_1};
\label{hdhdhd_4}
\end{eqnarray}
the general expressions for $ h_1,h_2 \text{ and } ~h_3 $ are given in the appendix \ref{app-Phase4} (see Eq. \ref{h1},~ \ref{h2} and \ref{h3}). 
Now imposing the condition (\ref{flux_11}), i.e. , equating (\ref{ldldhd_2J})(a) and (\ref{ldldhd_2J})(b), we get\\
\begin{eqnarray}
\frac{\beta_{eff1}(1-\beta_{eff1})}{1+\beta_{eff1}(\ell-1)}=\frac{\beta_{eff2}(W_s-\beta_{eff2})}{W_s+\beta_{eff2}(\ell-1)}, 
\label{hdhdhd_6}
\end{eqnarray}
solving which for the unknown $\beta_{eff1}$ we get,\\
\begin{align}
\beta_{eff1}=\frac{-q_1-\sqrt{q_1^2-4q_2q_3}}{2 q_3};
\label{hdhdhd_7}
\end{align}
the general expressions of $ q_1, q_2 \text{ and} ~q_3 $ are given in appendix \ref{app-Phase4} 
(see Eq. \ref{q1_1},~\ref{q2_1} and~ \ref{q3_1}).
Substituting Eqn.(\ref{hdhdhd_4}) into (\ref{hdhdhd_6}), we get the 
expression for $\beta_{eff1}$, in terms of $\beta$, $W_s$ and $\ell$. 
Thus, the expressions for canonical and non-canonical flux for HD/HD/HD phase are given by 
\begin{eqnarray}
J_1 &=& J_{seg1}=J_{seg2}=\frac{\beta_{eff2}(W_s-\beta_{eff2})}{W_s+\beta_{eff2}(\ell-1)}, \nonumber \\
{\rm and} \nonumber \\
J_{2} &=& \frac{\beta(1-\beta)}{1+\beta(\ell-1)}-\frac{\beta_{eff2}(W_s-\beta_{eff2})}{W_s+\beta_{eff2}(\ell-1)}.
\label{hdhdhd_8}
\end{eqnarray}
respectively.\\
\textbf{Phase 5: HD/MC/MC phase}
This phase can be specified by the following conditions,\\
\begin{eqnarray}
\beta_{eff1}  &<& \frac{1}{\sqrt{{\ell}}+1} , ~~~~ \beta_{eff1} < \alpha_1,  \nonumber \\
\beta_{eff2} &>& \frac{W_s}{\sqrt{{\ell}}+1} , ~~~~\alpha_{eff2}  >  \frac{W_s}{\sqrt{{\ell}}+1},  \nonumber \\
\beta & > & \frac{1}{\sqrt{{\ell}}+1} , ~~~~\alpha_{eff3}  > \frac{1}{\sqrt{{\ell}}+1}.
\label{hdmcmc_1}
\end{eqnarray}
The fluxes of the rods that follow from equations (\ref{hdp_2})(a) and (\ref{mcp_2})(a), are given by
\begin{eqnarray}
J_{seg1} &= \dfrac{\beta_{eff1}(1-\beta_{eff1})}{1+\beta_{eff1}(\ell-1)} ,~~(a)\nonumber \\
 J_{seg2} & = \dfrac{W_s}{(\sqrt{{\ell}}+1)^{2}} ,~~(b) \nonumber \\
 J_{seg3} &=   \dfrac{1}{(\sqrt{{\ell}}+1)^{2}},~~(c) \nonumber \\
 \label{hdmcmc_2}
\end{eqnarray}
while, from the equations (\ref{hdp_2})(b) and (\ref{mcp_2})(b), the corresponding coverage densities are given by
\begin{eqnarray}
\rho_{c1}&=&1-\beta_{eff1}, ~~(a)\nonumber \\
 \rho_{c2}&=&\frac{\sqrt{{\ell}}}{\sqrt{{\ell}}+1}, ~~(b)\nonumber \\
  \rho_{c3}&=&  \frac{\sqrt{{\ell}}}{\sqrt{{\ell}}+1}.~~(c)\nonumber \\
 \label{hdmcmc_3}
\end{eqnarray}

Now imposing the condition (\ref{flux_11}), i.e., equating  (\ref{hdmcmc_2})(a) and  (\ref{hdmcmc_2})(b), we get\\
\begin{align}
\frac{\beta_{eff1}(1-\beta_{eff1})}{1+\beta_{eff1}(\ell-1)}=\frac{W_s}{(\sqrt{{\ell}}+1)^{2}} ,
\label{hdmcmc_4}
\end{align}
solving which for $\beta_{eff1}$ we get
\begin{equation}
\beta_{eff1}=\frac{-k_1-\sqrt{(k_1)^2+4 k_2}}{2 k_3},
\label{hdmcmc_5}
\end{equation}
where,
\begin{eqnarray}
k_1 &=& 1+W_s-W_s\ell +\ell+2 \sqrt{\ell}, \nonumber \\
k_2 &=& (-\ell-2 \sqrt{\ell}-1) W_s, \nonumber \\
k_3 &=& (-\ell-2 \sqrt{\ell}-1).
\end{eqnarray}
From condition (\ref{hdmcmc_1}) it follows that the HD/MC/MC phase exists on the $\alpha_{1}-\beta$ plane in the region where
\begin{eqnarray}
\alpha_1  &>&\frac{-k_1-\sqrt{(k_1)^2+4 k_2}}{2 k_3}, \nonumber \\
\beta &>& \frac{1}{\sqrt{{\ell}}+1}.
\label{hdmcmc_6}
\end{eqnarray}
The corresponding canonical and non-canonical flux in the HD/MC/MC phase are
\begin{eqnarray}
J_1& =J_{seg1}=J_{seg2}=\frac{W_s}{(\sqrt{{\ell}}+1)^{2}}, \nonumber \\ 
{\rm and} \nonumber \\
J_2 &=\frac{1}{(\sqrt{{\ell}}+1)^{2}}-\frac{W_s}{(\sqrt{{\ell}}+1)^{2}}.
\label{hdmcmc_8}
\end{eqnarray}
respectively.

\noindent \textbf{Phase 6: HD/MC/LD phase}\\

This phase can be specified by the following conditions:\\
\begin{eqnarray}
\beta_{eff1}  &<& \frac{1}{\sqrt{{\ell}}+1} , ~~~~ \beta_{eff1} < \alpha_1,  \nonumber \\
\beta_{eff2} &>& \frac{W_s}{\sqrt{{\ell}}+1} , ~~~~\alpha_{eff2}  > \frac{W_s}{\sqrt{{\ell}}+1},  \nonumber \\
\alpha_{eff3}  &<& \frac{1}{\sqrt{{\ell}}+1} , ~~~~ \alpha_{eff3} < \beta.
\label{hdmcld_1}
\end{eqnarray}
Under steady state condition the fluxes of the rods that follow from equations (\ref{hdp_2} a), (\ref{mcp_2} a) and (\ref{ldp_2} a), are given by\\
\begin{eqnarray}
J_{seg1} &=& \frac{\beta_{eff1}(1-\beta_{eff1})}{1+\beta_{eff1}(\ell-1)} ,~~(a)\nonumber \\
 J_{seg2} &=&\frac{W_s}{(\sqrt{{\ell}}+1)^{2}} ,~~(b)\nonumber \\
 J_{seg3} &=&  \frac{\alpha_{eff3}(1-\alpha_{eff3})}{1+\alpha_{eff3}(\ell-1)},~~(c)\nonumber \\
 \label{hdmcld_2}
\end{eqnarray}
while, from the equations (\ref{hdp_2} b), (\ref{mcp_2} b) and (\ref{ldp_2} b), the corresponding coverage densities are given by,\\
\begin{eqnarray}
\rho_{c1} &=& 1-\beta_{eff1}, ~~(a)\nonumber \\
 \rho_{c2} &=& \frac{\sqrt{{\ell}}}{\sqrt{{\ell}}+1}, ~~(b)\nonumber \\
 \rho_{c3} &=& \frac{\alpha_{eff3}\ell}{1+\alpha_{eff3}(\ell-1)}.~~(c)\nonumber \\
 \label{hdmcld_3}
\end{eqnarray}

Now imposing the condition (\ref{flux_11}), i.e., equating  (\ref{hdmcld_2})(a) and  (\ref{hdmcld_2})(b), we get 
a quadratic equation for $\beta_{eff1}$ whose solution is given by (\ref{hdmcmc_5}).  Next, using Eq. (\ref{flux_22}), i.e., equating the fluxes of segment \RN{2} and \RN{3} we get
\begin{eqnarray}
\frac{\alpha_{eff3}(1-\alpha_{eff3})}{1+\alpha_{eff3}(\ell-1)}=\frac{W_s}{(\sqrt{{\ell}}+1)^{2}}+\alpha_{IE}P(\underbrace{0, \dots,0}_{\ell}) , \nonumber \\
\label{hdmcld_7}
\end{eqnarray}
for HD/MC/LD phase (see Eq. (\ref{ld_1}) in the Appendix)\\
\begin{equation}
P(\underbrace{0, \dots,0}_{\ell})=\dfrac{2(1-\alpha_{eff3})}{2(1-\alpha_{eff3})(1+\alpha_{eff3}^\prime) +\alpha_{eff3}\ell(2+\alpha_{eff3}^\prime)},
\label{hdmcld_8}
\end{equation}
where, $ \alpha_{eff3}^\prime = \alpha_{eff3}(\ell-1) $.\\
Since, we are taking the special site $ i_s $, far from the boundary between segment \RN{2} and \RN{3}, $ \alpha_{NIE} $ is not affected by $ \alpha_{IE} $. Therefore, to simplify the calculation, we first solve equation (\ref{hdmcld_7}) for $ \alpha_{IE}=0  $ getting the expression for $ \alpha_{eff3}=\alpha_{NIE}$, in terms of $W_s$ and $\ell$, to be
\begin{eqnarray}
\alpha_{NIE}=\frac{-k_1-\sqrt{(k_1)^2+4 k_2}}{2 k_3},
\label{hdmcld_9}
\end{eqnarray}
and hence in the general case of of $\alpha_{IE}\neq 0$,
$\alpha_{eff3}=\alpha_{NIE} +\alpha_{IE}$, i.e., 
\begin{eqnarray}
\alpha_{eff3}=\frac{-k_1-\sqrt{(k_1)^2+4 k_2}}{2 k_3} +\alpha_{IE}.
\label{hdmcld_11}
\end{eqnarray}

From condition (\ref{hdmcld_1}), we find that the segment \RN{3} would be in LD phase for
\begin{eqnarray}
\alpha_{NIE}+\alpha_{IE} < \frac{1}{(\sqrt{\ell}+1)}.
\label{hdmcld_12a}
\end{eqnarray} 
But, since, the Eqn. (\ref{hdmcld_12a}) is independent of $ \alpha_1 $ this phase cannot be displayed 
on the $\alpha_1 $ - $ \beta$ plane in case of signal-independent initiation. However, in the case of signal-dependent initiation
\begin{eqnarray}
\alpha_{IE}=\alpha- \alpha_1 \\ \nonumber
\label{hdmcld_12b}
\end{eqnarray} 
the HD/MC/LD phase appears on the $ \alpha_1 $ - $ \beta $ plane in the region where
$(\alpha_{NIE}+\alpha-\alpha_1) < 1/(\sqrt{\ell}+1)$,  
i.e.,
\begin{eqnarray}
\alpha_1 &>& \alpha + \frac{-k_1-\sqrt{(k_1)^2+4 k_2}}{2 k_3} -\frac{1}{(\sqrt{\ell}+1)}, \nonumber \\
\beta &>&  \frac{-k_1-\sqrt{(k_1)^2+4 k_2}}{2 k_3} +\alpha -\alpha_1.
\label{hdmcld_12d}
\end{eqnarray}

For a given $\alpha_1$, when 
$\alpha_1  > \alpha+ \alpha_{NIE}-1/(\sqrt{\ell}+1) $, 
if $\beta$ is increased so that it exceeds $\alpha_{eff3}$  
the segment \RN{3} transforms to LD phase from HD phase and the 
composite phase of the system makes a transition from HD/MC/HD 
to HD/MC/LD.
Similarly, For a given $\beta$, when $\beta > 1/(\sqrt{\ell}+1)$, 
decrease of $\alpha_1$ or increase of $\alpha_{IE}$ can result 
in a value of $\alpha_{eff3}$ that just exceeds $ 1/(\sqrt{\ell}+1)$.  
At this point a transition of segment \RN{3} from LD phase 
to MC phase takes place and, consequently, the system, as a whole, 
exhibits a transition from composite phase HD/MC/LD to HD/MC/MC.

Finally, the canonical and non-canonical fluxes in the HD/MC/LD phase are 
\begin{eqnarray}
J_1& =J_{seg1}=J_{seg2}=\frac{W_s}{(\sqrt{{\ell}}+1)^{2}}, \nonumber \\ 
J_2 &=\frac{\alpha_{eff3}(1-\alpha_{eff3})}{1+\alpha_{eff3}(\ell-1)}-\frac{W_s}{(\sqrt{{\ell}}+1)^{2}}.
\label{hdmcld_13}
\end{eqnarray}

\begin{figure}[htp]
  \includegraphics[width=.4\textwidth]{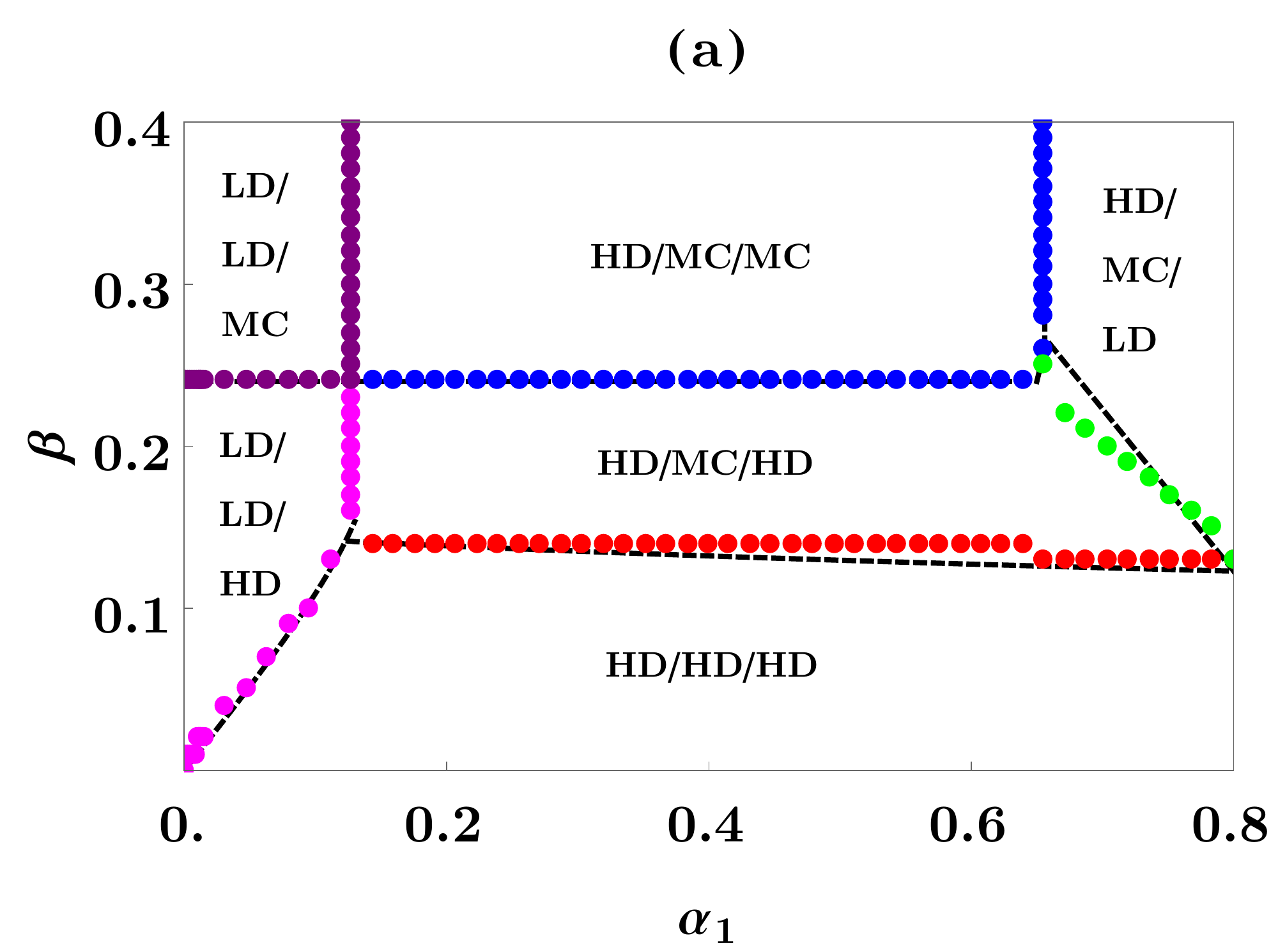} \\  
  \includegraphics[width=.4\textwidth]{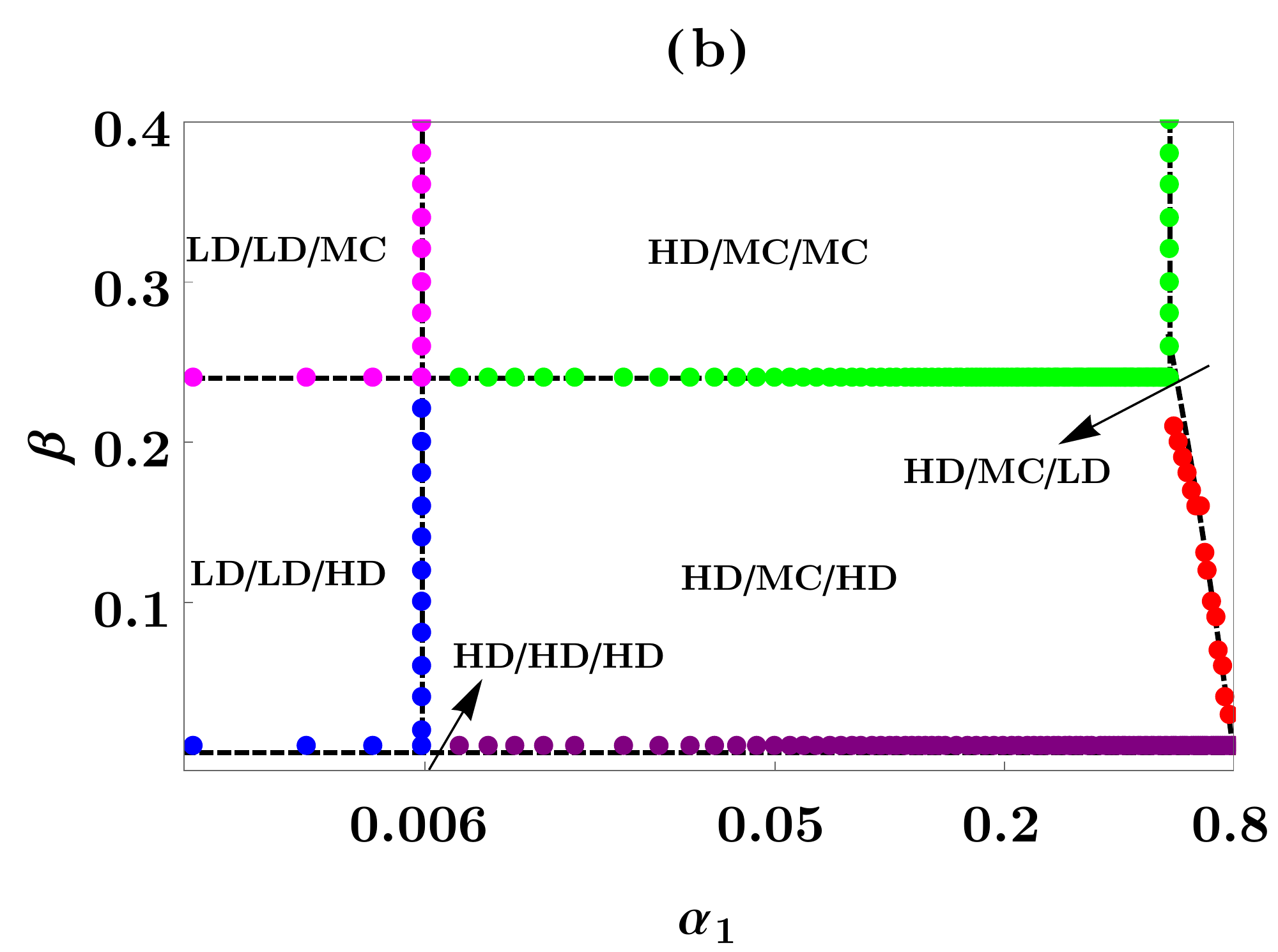} 
\caption{(Color online) Phase diagrams of the model in the $\alpha_{1}-\beta$ plane, 
in the case of signal-dependent initiation, are plotted for 
(a) $ \gamma=0.9 $, 
and (b) $ \gamma=0.1$. The theoretical prediction obtained under MFA are 
drawn by dashed curves while the numerical data obtained from Monte-Carlo 
simulations are shown by discrete symbols. The numerical values of the 
relevant parameters used here are $L+\ell-1=1200+\ell-1$, $\ell=10$, 
$i_s=530,n=3,m=5$, $ \alpha=0.8 $ and $ W=1 $.}
\label{phase_1_dep}
\end{figure}
\begin{figure}[htp]
  \includegraphics[width=.3\textwidth]{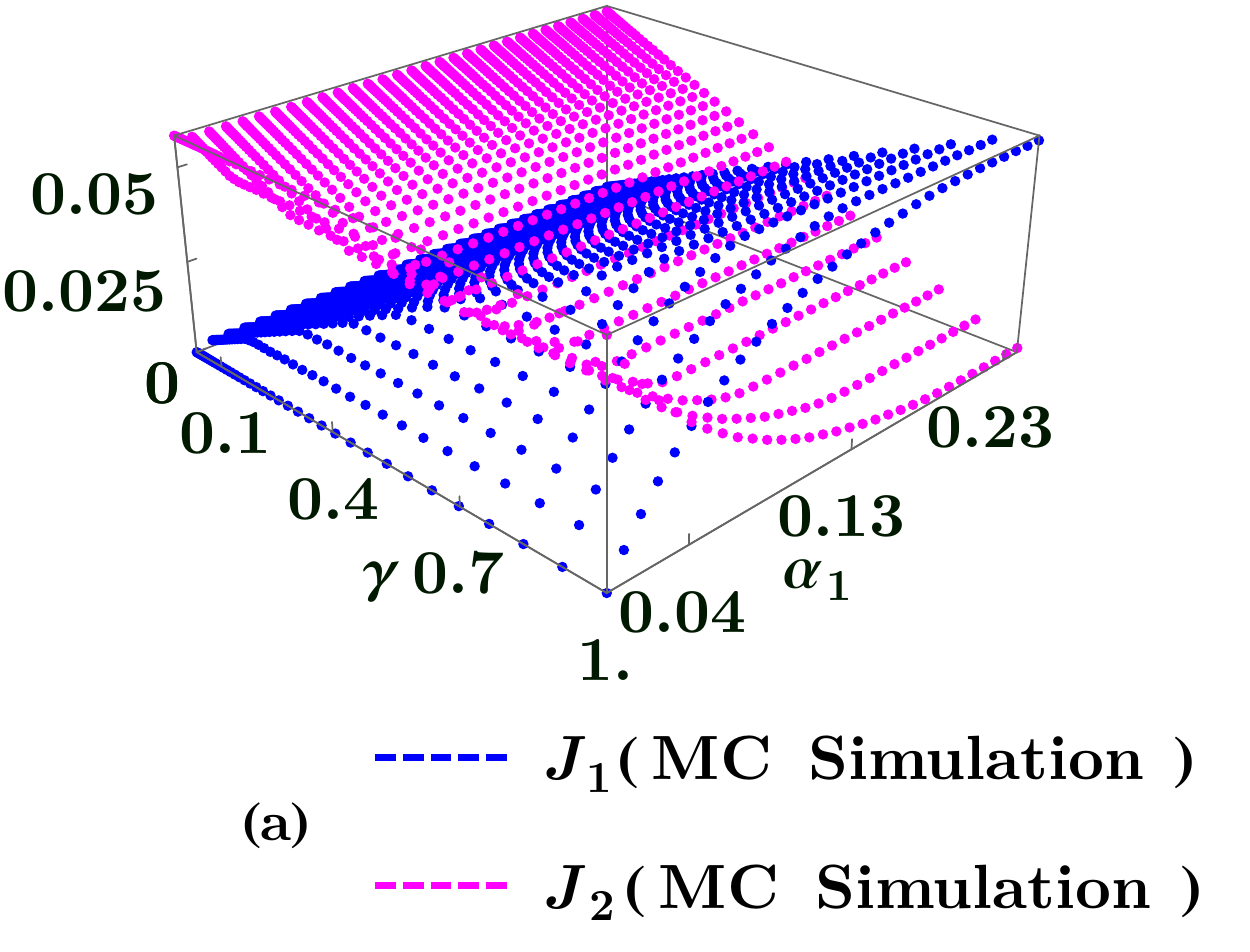} 
  \includegraphics[width=.3\textwidth]{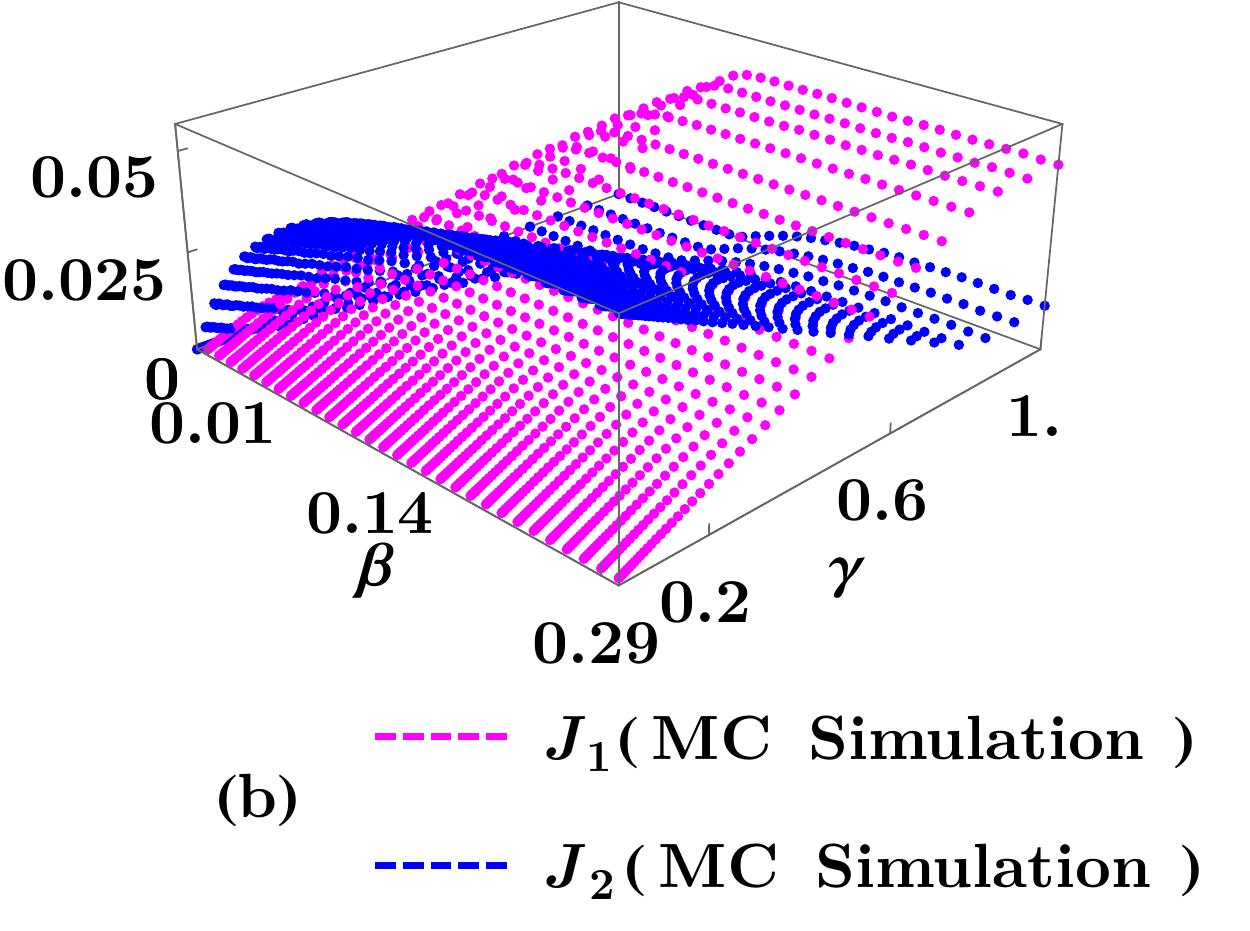}    
  \includegraphics[width=.3\textwidth]{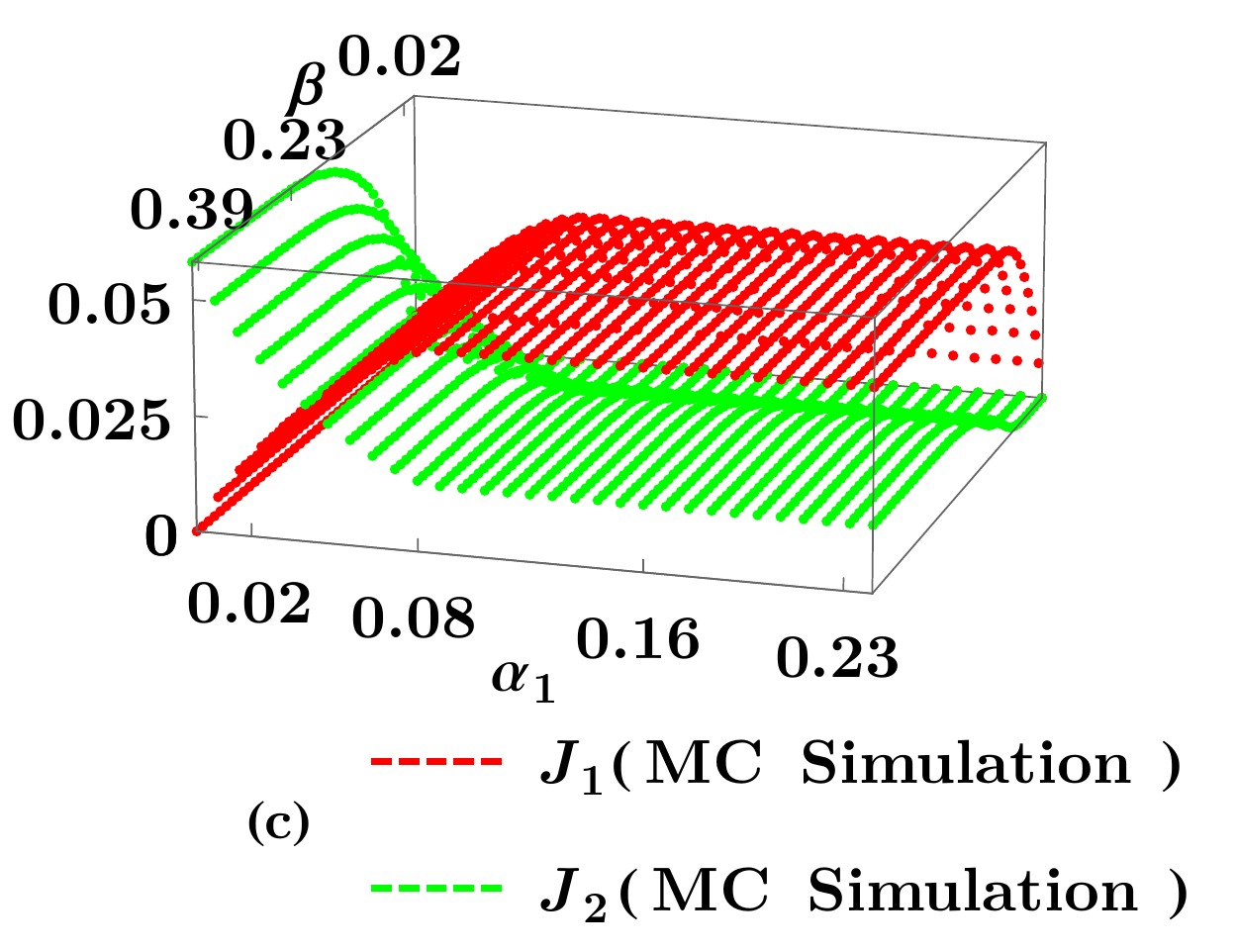} 
    \caption{(Color online) 3D plots of the fluxes $J_{1}$ and $J_{2}$ of the $rod_1$ and 
$rod_2$, respectively, against (a) $\alpha_{1}$ and $\gamma$ (for constant $ \beta=0.8 $), (b) $\gamma$ and $\beta$ (for constant $ s=0.1 $), (c) $\alpha_{1}$ and $\beta$ (for constant $ \gamma=0.47 $), respectively, in the case of signal-dependent initiation. The numerical values of the other relevant parameters used in this figure are $ L+\ell-1=1200+\ell-1 $, $ \ell=10 $, $ i_s=512,n=1,m=7 $, $ W=1 $ and $ \alpha=0.8 $.}
  \label{flux_3d}
\end{figure}
\begin{figure}[t] 
\begin{center}
\includegraphics[width=0.75\columnwidth]{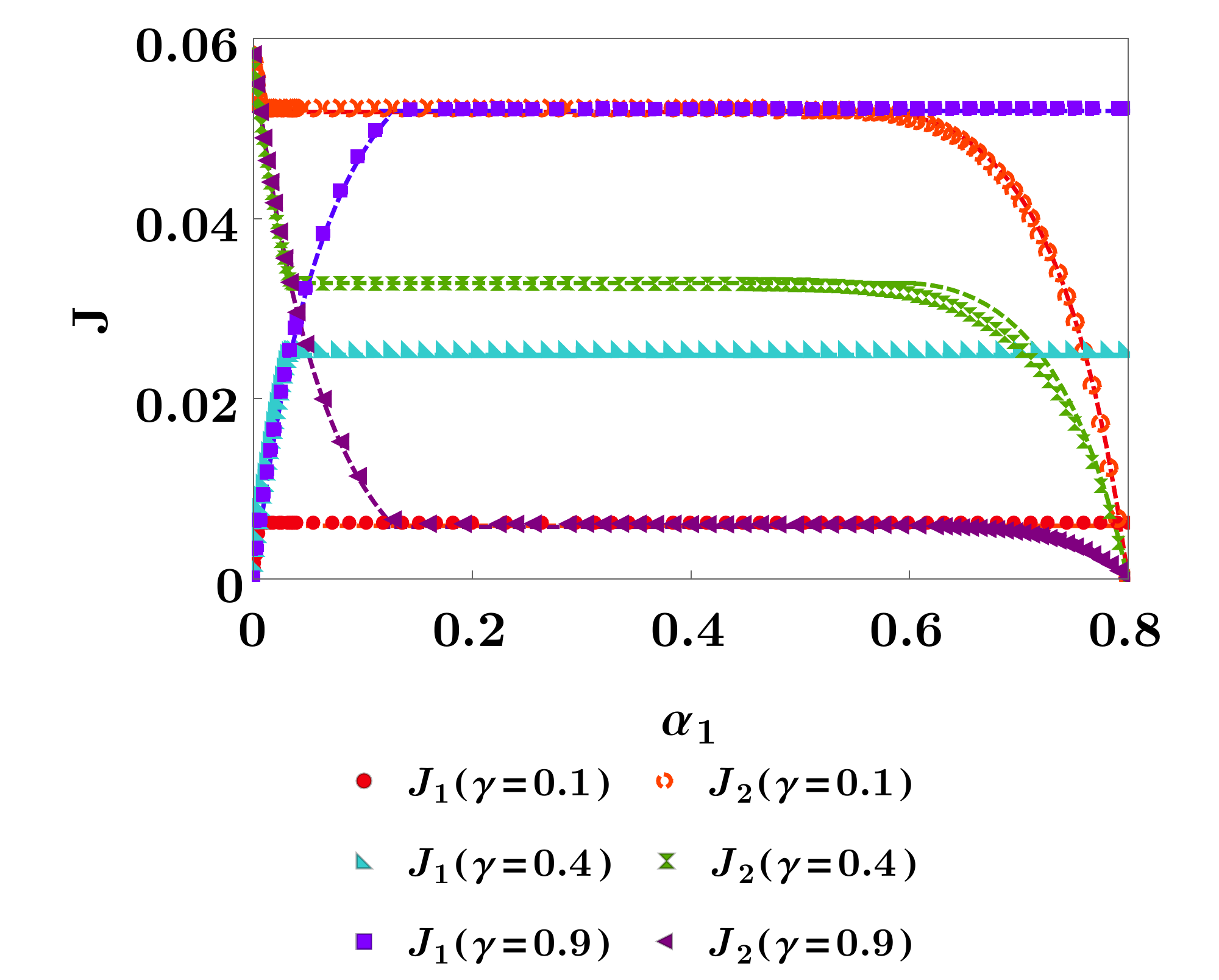} 
\end{center} 
\caption{(Color online) Two-dimensional cross sections of Fig.(\ref{flux_3d}(a)) corresponding to three different values of $\gamma$ are plotted. 
Thus, fluxes $J_{1}$ and $J_{2}$ are plotted against $\alpha_{1}$ for 
three different values of $\gamma$. 
The theoretical predictions obtained under MFA are drawn with continuous 
curves and the numerical data obtained from Monte-Carlo simulations are 
shown with discrete symbols. The numerical values of the other relevant 
parameters used in the figure are $ L+\ell-1=1200+\ell-1$, $ \ell=10 $,  
$i_s=512,n=1,m=7$, $W=1$, $\alpha=0.8$ and $\beta=0.8$.}
\label{flux_2d_1}
\end{figure}
\begin{figure}[t] 
\begin{center}
\includegraphics[width=0.75\columnwidth]{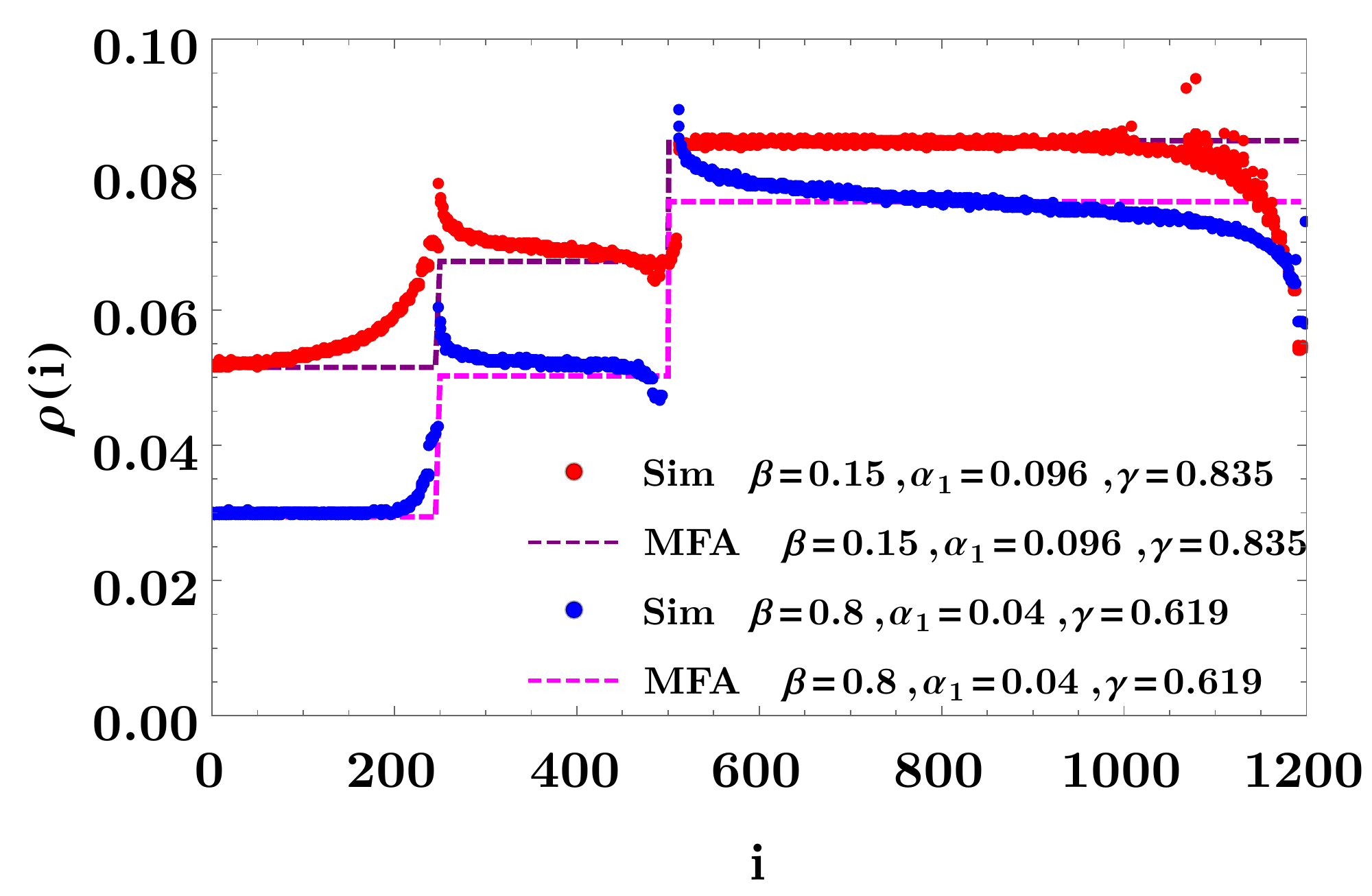}  
\end{center} 
\caption{(Color online) Density profiles of rods ($ \rho(i)=P_1(i)+P_2(i)$) in the case 
of signal-dependent initiation are plotted 
for two sets of values of $\beta, \alpha_{1}$ and $\gamma$. 
The theoretical prediction obtained under MFA are drawn 
by continues curves and numerical data obtained from Monte-Carlo simulations 
are shown with discrete symbols. The numerical values of the other relevant 
parameters used in this figure are $ L+\ell-1=1200+\ell-1 $, $ \ell=10 $, $ i_s=512,n=1,m=7 $, $ W=1 $ and $ \alpha=0.8 $. The 
blue curve corresponds to the LD/LD/MC phase where as the red curve 
corresponds to the LD/LD/HD phase.} 
\label{density_profile}
\end{figure}
\begin{figure}[htp]
 \includegraphics[width=.4\textwidth]{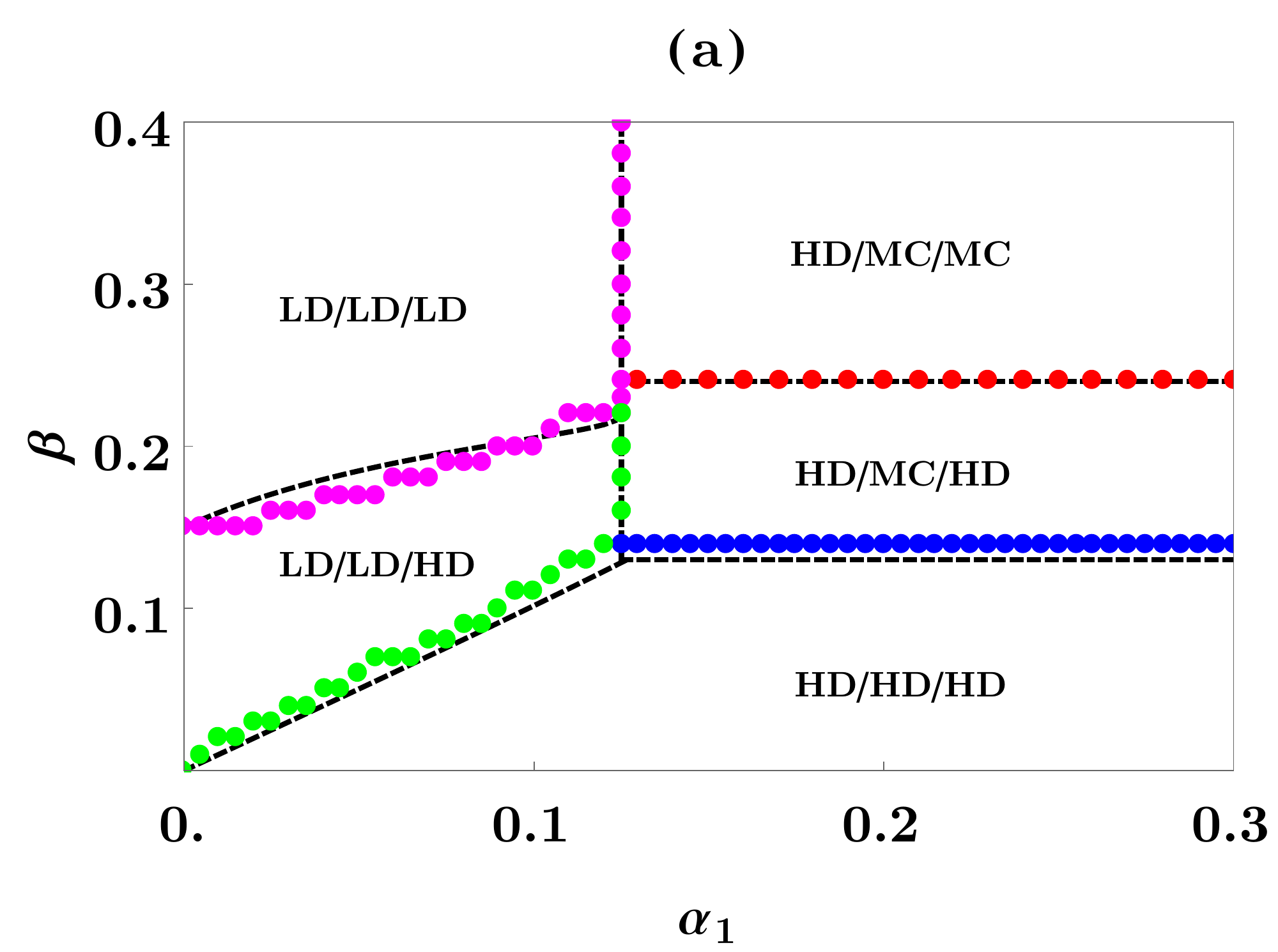} 
  \includegraphics[width=.4\textwidth]{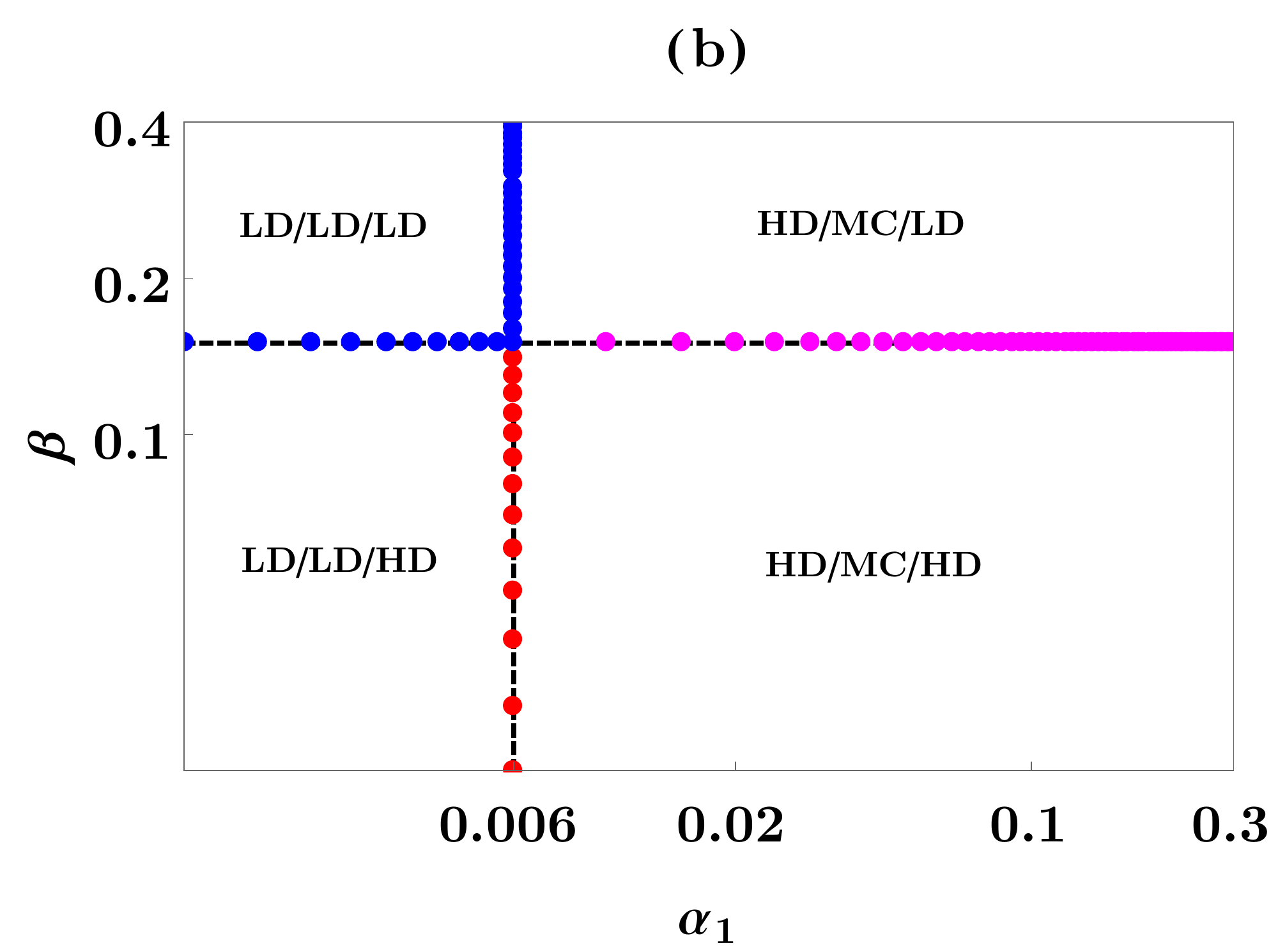}
\caption{(Color online) Phase diagrams of the model in the case of signal-independent initiation are plotted keeping $ \gamma=0.9 $ and $ \alpha_{IE}=0.15 $, fixed in (a) and $ \gamma=0.1 $ and $ \alpha_{IE}=0.15 $ fixed in (b).  
The theoretical prediction obtained under MFA are drawn by continues curves and numerical data obtained from Monte-Carlo simulations are shown with discrete symbols. The numerical values of the other relevant parameters used in this figure are $L+\ell-1=1200+\ell-1$, $\ell=10$, $i_s=512,n=1,m=7$, and $W=1$.}
\label{phase_1_ind}
\end{figure}

\begin{figure}[htp]
  \includegraphics[width=.4\textwidth]{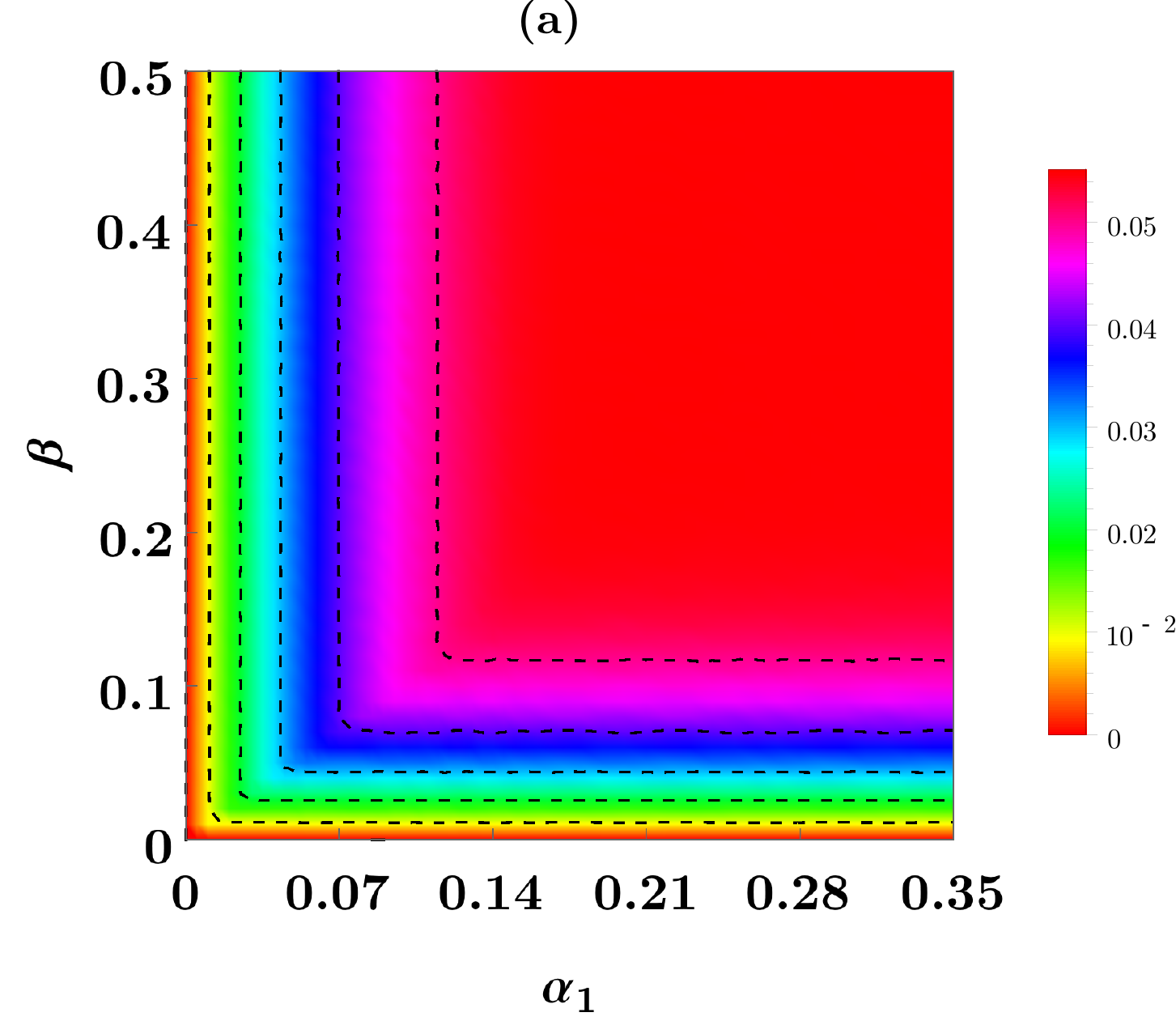} 
  \includegraphics[width=.4\textwidth]{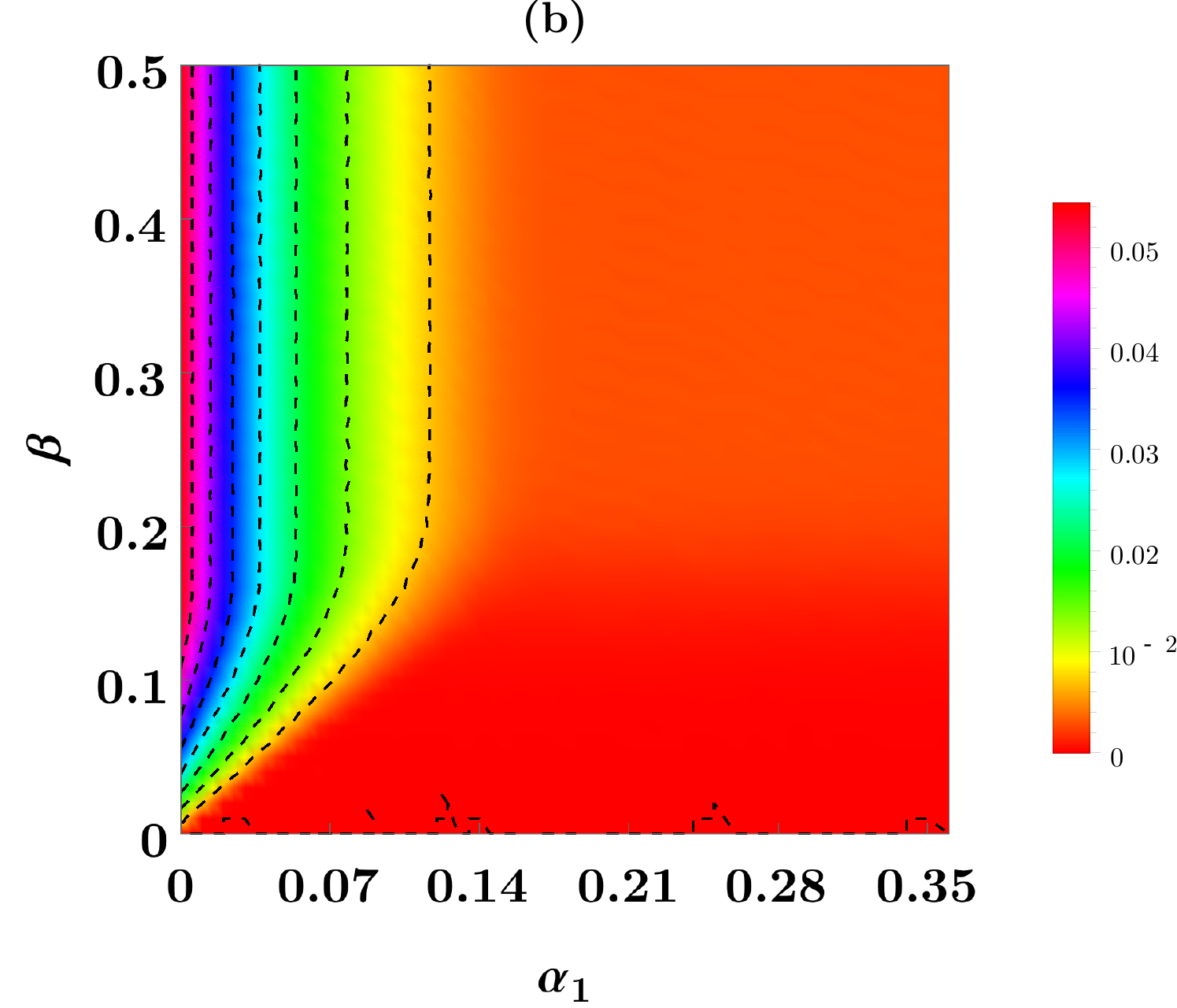} 
\caption{(Color online) Contour plots of the fluxes $J_1$ and $J_{2}$ in the $\alpha_{1}-\beta$ plane shown in (a) and (b) are obtained from Monte Carlo simulations of the model in the case of signal-independent initiation. The numerical values of the other relevant parameters are $ L+\ell-1=1200+\ell-1$, $\ell=10$, $ i_s=512,n=1,m=7 $, $ W=1 $,$ \alpha_{IE}=0.35 $ and $ \gamma=0.74 $.}
\label{contour_plot_a}
\end{figure}

\begin{figure}[htp]
  \includegraphics[width=.4\textwidth]{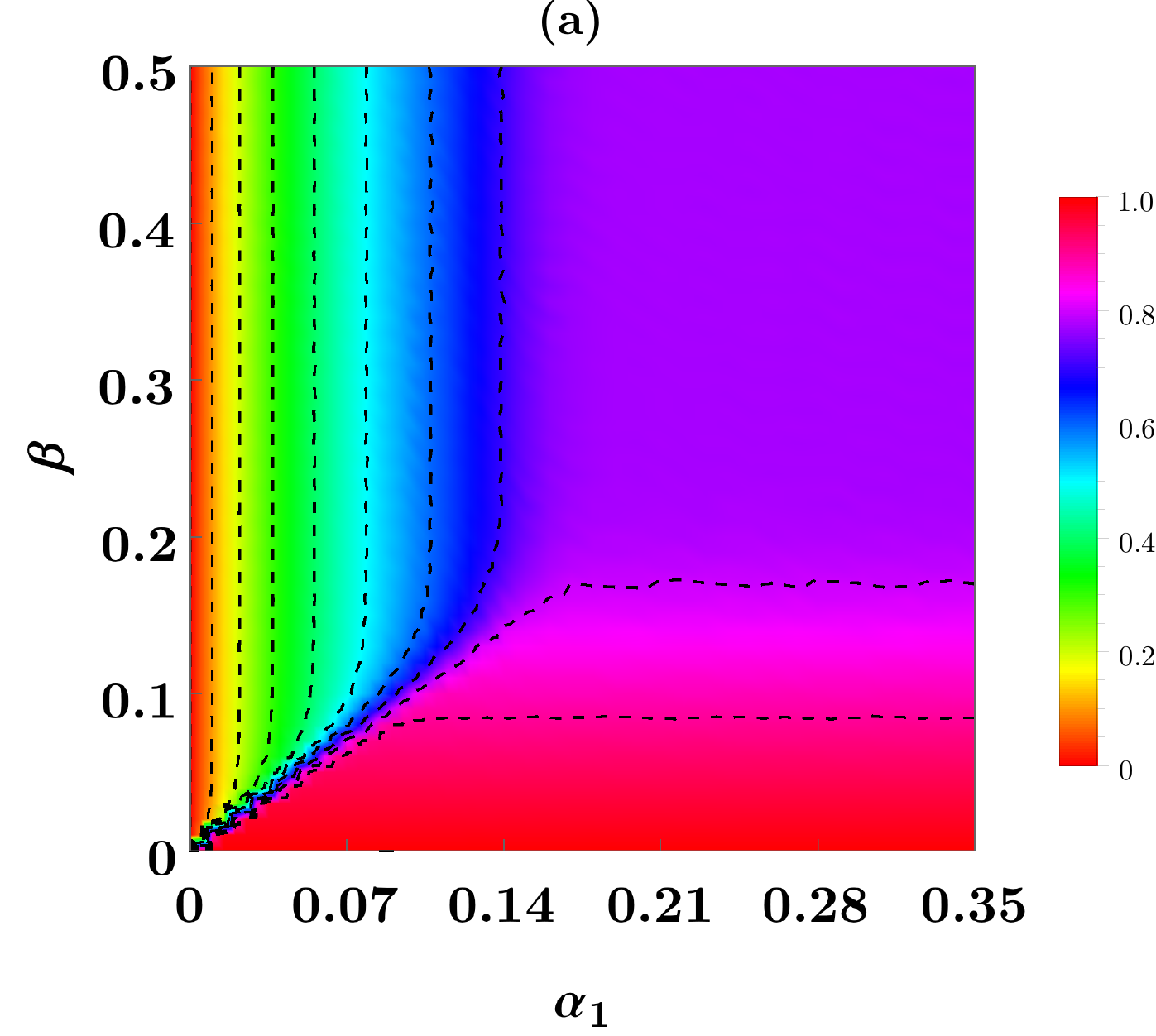} 
  \includegraphics[width=.4\textwidth]{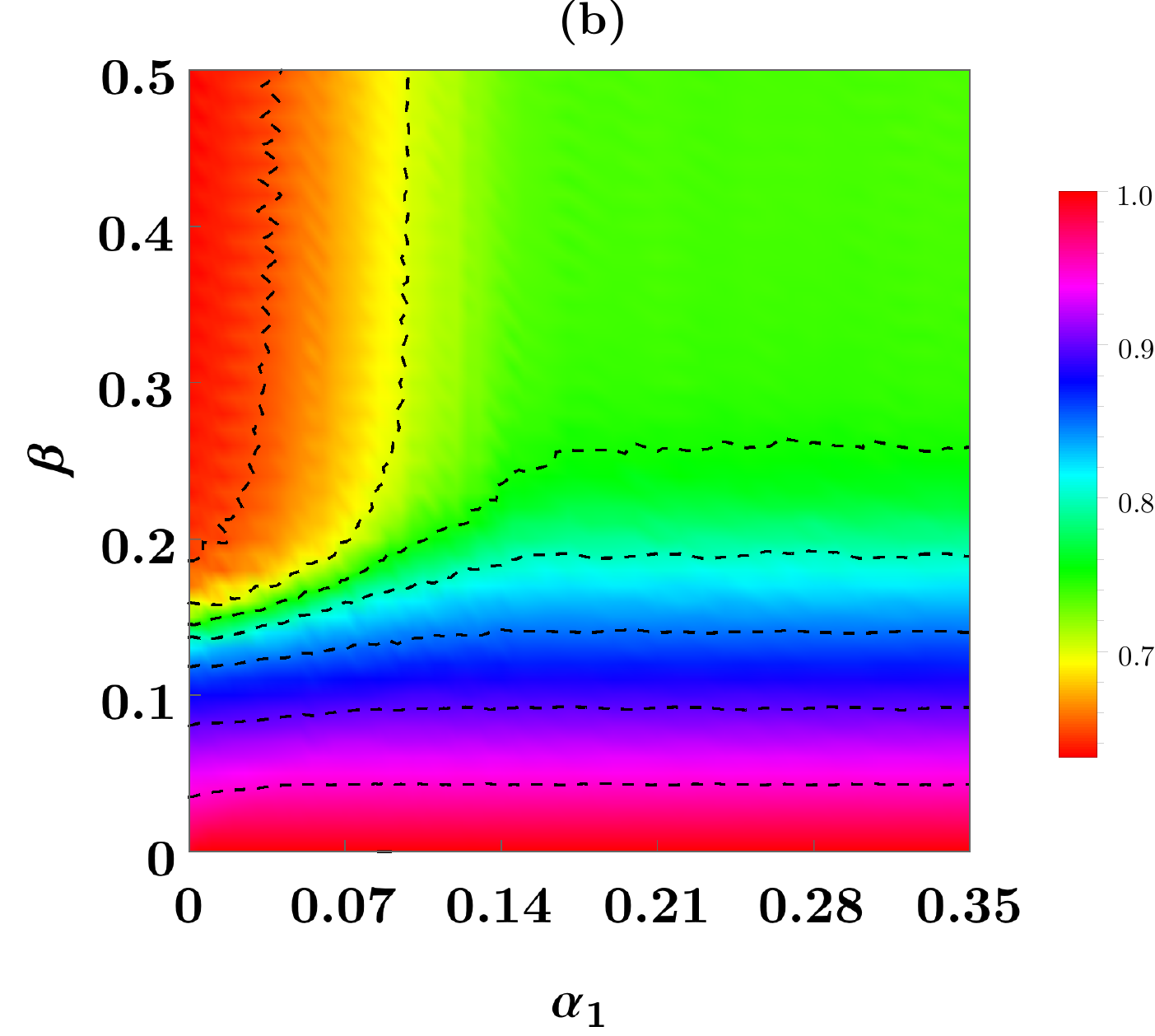} 
\caption{(Color online) Contour plots of the coverage densities in the segments \RN{1} and \RN{3}  in the $\alpha_{1}-\beta$ plane shown in (a) and (b) are obtained from Monte Carlo simulations of the model in the case of signal-independent initiation. The numerical values of the other relevant parameters are $ L+\ell-1=1200+\ell-1$, $\ell=10$, $i_s=512,n=1,m=7 $, $ W=1 $,$ \alpha_{IE}=0.35 $ and $ \gamma=0.74 $.}
\label{contour_plot_b}
\end{figure}

\begin{figure}[t] 
\begin{center}
\includegraphics[width=0.75\columnwidth]{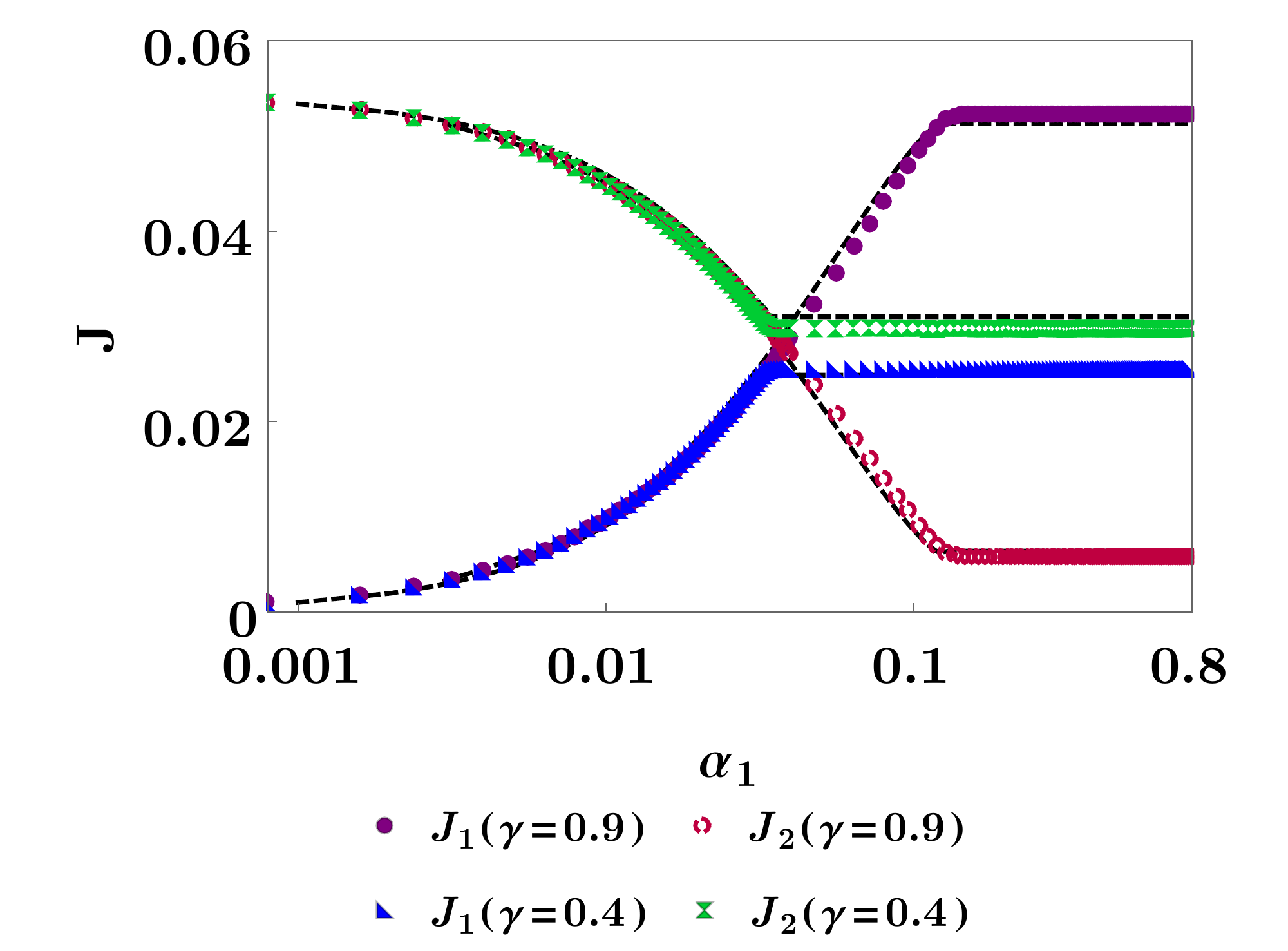}  
\end{center} 
\caption{(Color online) Fluxes $J_{1}$ and $J_{2}$ in the case of signal-independent initiation are plotted against $\alpha_1$, for two different values of $\gamma$ and for a fixed $ \beta$ . 
The theoretical prediction obtained under MFA are drawn by continues curves and numerical data obtained from Monte-Carlo simulations are shown with discrete symbols. The numerical values of the other relevant parameters used in this figure are $ L+\ell-1=1200+\ell-1 $, $ \ell=10 $, $ i_s=512,n=1,m=7 $, $ W=1 $ and $ \beta=0.8 $.} 
\label{flux_2d_3}
\end{figure}
\begin{figure}[t] 
\begin{center}
\includegraphics[width=0.75\columnwidth]{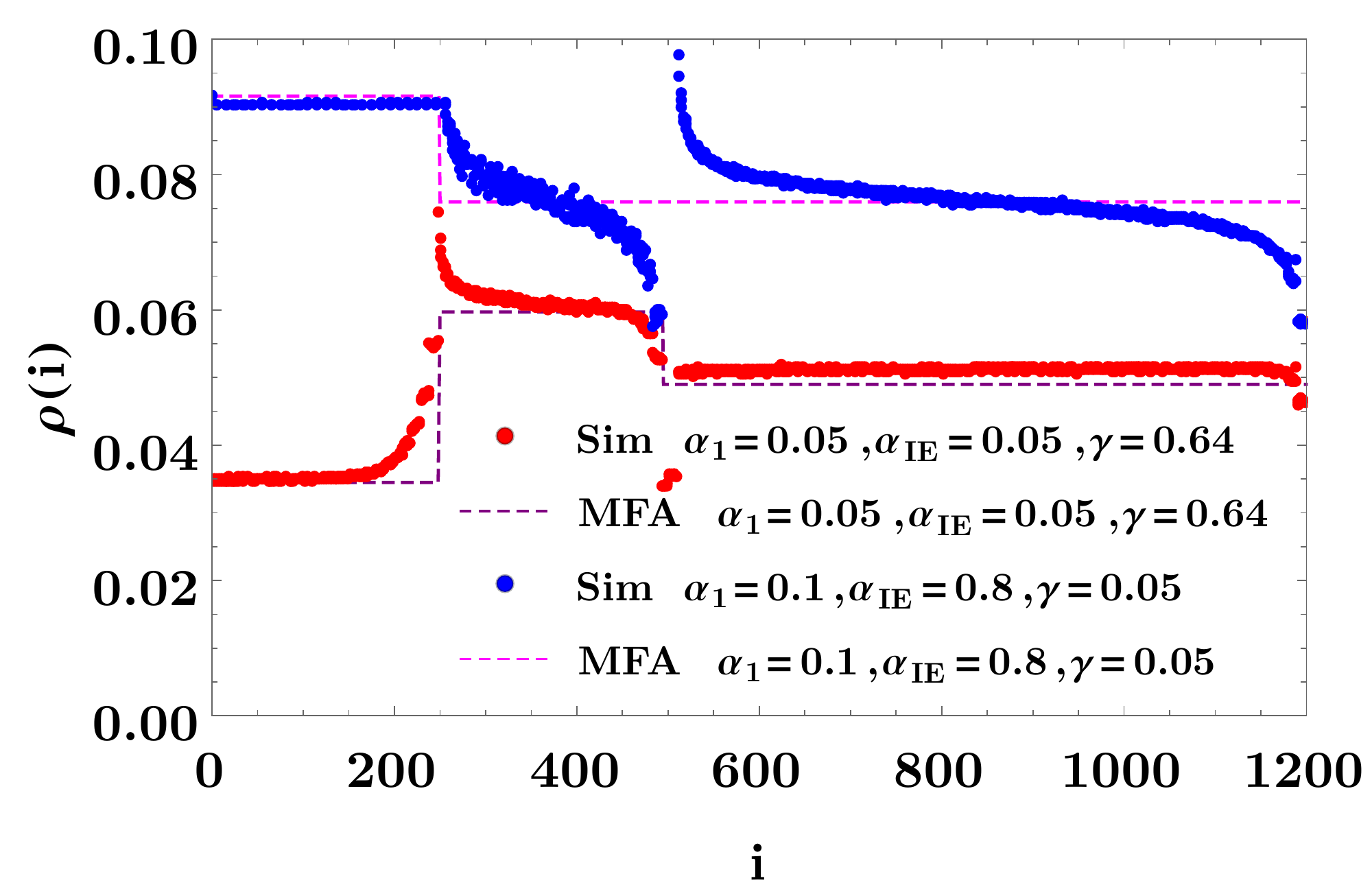}  
\end{center} 
\caption{(Color online) Density profiles of rods ($ \rho(i)=P_1(i)+P_2(i) $) are plotted 
for two sets of values of the parameters $\alpha_{1}, \alpha_{IE}$ and $\gamma$. The theoretical prediction obtained under MFA are drawn by continues curves and numerical data obtained from Monte-Carlo simulations are shown with discrete symbols. The numerical values of the other relevant parameters used in this figure are $ L+\ell-1=1200+\ell-1 $, $ \ell=10 $, $ i_s=512,n=1,m=7 $, $ \beta=0.8 $ and $ W=1 $.}
\label{density_profile_2}
\end{figure}
\textbf{Phase 7: HD/MC/HD phase}
This phase can be specified by the following conditions,\\
\begin{eqnarray}
\beta_{eff1}  &<& \frac{1}{\sqrt{{\ell}}+1} , ~~~~ \beta_{eff1} < \alpha_1,  \nonumber \\
\beta_{eff2} &>& \frac{W_s}{\sqrt{{\ell}}+1} , ~~~~\alpha_{eff2}  > \frac{W_s}{\sqrt{{\ell}}+1},  \nonumber \\
\beta  &<& \frac{1}{\sqrt{{\ell}}+1} , ~~~~  \beta<\alpha_{eff3}.
\label{hdmchd_1}
\end{eqnarray}
Under stationary state condition the fluxes of the rods that follow from equations (\ref{hdp_2} a) and \ref{mcp_2}) a), are given by\\
\begin{eqnarray}
J_{seg1} &=& \frac{\beta_{eff1}(1-\beta_{eff1})}{1+\beta_{eff1}(\ell-1)} ,~~(a)\nonumber \\
 J_{seg2} &=& \frac{W_s}{(\sqrt{{\ell}}+1)^{2}} ,~~(b)\nonumber \\
 J_{seg3} &=&  \frac{\beta(1-\beta)}{1+\beta(\ell-1)},~~(c)\nonumber \\
 \label{hdmchd_2}
\end{eqnarray}
while, from the equations (\ref{hdp_2} b) and (\ref{mcp_2} b), the corresponding coverage densities are given by,\\
\begin{eqnarray}
\rho_{c1}  &=& 1-\beta_{eff1}, ~~(a)\nonumber \\
 \rho_{c2} &=& \frac{\sqrt{{\ell}}}{\sqrt{{\ell}}+1}, ~~(b)\nonumber \\
 \rho_{c3} &=& 1-\beta.~~(c)\nonumber \\
 \label{hdmchd_3}
\end{eqnarray}

Now imposing the condition (\ref{flux_11}), i.e., equating  (\ref{hdmchd_2})(a) and  (\ref{hdmchd_2})(b), we get 
a quadratic equation for $\beta_{eff1}$ whose solution is given by (\ref{hdmcmc_5}). Imposing Eq. (\ref{flux_22}) we now get 
\begin{eqnarray}
\frac{\beta(1-\beta)}{1+\beta(\ell-1)}=\frac{W_s}{(\sqrt{{\ell}}+1)^{2}}+\alpha_{IE}P(\underbrace{0, \dots,0}_{\ell}) ,
\label{hdmchd_7}
\end{eqnarray}
where substitution of 
\begin{equation}
P(\underbrace{0, \dots,0}_{\ell})=\dfrac{2\beta}{2\beta(1+\beta^\prime) +(1-\beta)\ell(2+\beta^\prime)},
\label{hdmchd_8}
\end{equation}
for HD/MC/HD phase (see the derivation of Eq. (\ref{hd_1}) in the Appendix),
with $\beta^\prime = (1-\beta)(\ell-1)$, we get an equation whose solution 
yields the expression for $ \beta $ in terms of $ W_s $, $ \alpha_{IE} $ and $ \ell $
.\\

Solving this quartic equation of $ \beta $, we get the expression of 
$\beta$ in terms of $W_s$, $\ell$ and $\alpha_{IE}$. Out of the four 
solutions of the quartic equation, two are the negative and, therefore, 
not physically admissible. Out of two positive solutions we have chosen 
the one with the negative sign to be the physical solution because, as 
restricted by (\ref{hdmchd_1}), $\beta$ must be less than 
$1/(\sqrt{{\ell}}+1)$. The general expression is too complicated to be 
reproduced here. Instead, the solution for a specific set of parameter 
values, namely, $ W_s=0.9$ and $\ell=10$, is given in the 
appendix \ref{app-Phase7}.

From condition (\ref{hdmchd_1}) occurrence of the HD/MC/HD phase 
requires
\begin{eqnarray}
\alpha_1  &>& \frac{-k_1-\sqrt{(k_1)^2+4 k_2}}{2 k_3} \nonumber \\
\frac{1}{(\sqrt{\ell}+1)} &>& \beta >  \bigg(z_0-0.5 \sqrt{z_1+\frac{z_2}{z_3}+z_4} \nonumber \\
&-&0.5 \sqrt{z_5-\frac{z_6}{z_7}-z_8-\frac{z_9}{z_{10}}}\bigg) \nonumber \\
\label{hdmchd_10}
\end{eqnarray}
(As example, for the expressions of $z_{\mu}(\mu=0,1, \cdots 10$) see appendix \ref{app-Phase7} )
For any given $ \alpha_1 $ that satisfies the condition 
$\alpha_1  > (-k_1-\sqrt{(k_1)^2+4 k_2})/(2 k_3) $, 
the phase in the segment \RN{3} transforms from HD phase to MC phase 
when $\beta$ just exceeds the value $ 1/(\sqrt{\ell}+1)$ and the 
composite phase of the system makes a transition from HD/MC/HD to HD/MC/MC. 
On the other hand, as $\beta$ decreases, the phase in segment \RN{2} 
transforms from MC phase to HD phase at 
$\beta=  (z_0-0.5 \sqrt{z_1+(z_2/z_3)+z_4} - 0.5 \sqrt{z_5-(z_6/z_7)-z_8-(z_9/z_{10})})$
and the composite phase of the system makes a transition from HD/MC/HD 
to HD/HD/HD. The physical reason for this transition is that, below this 
critical value of $\beta$, the segment \RN{3} creates an effective 
bottleneck at the end of the segment \RN{2} for the rods moving forward 
in segment \RN{2}.


The canonical and non-canonical fluxes in the HD/MC/HD phase are\\
\begin{eqnarray}
J_1& =&J_{seg1}=J_{seg2}=\frac{W_s}{(\sqrt{{\ell}}+1)^{2}}, \nonumber \\ 
J_2 &=&\frac{\beta(1-\beta)}{1+\beta(\ell-1)}-\frac{W_s}{(\sqrt{{\ell}}+1)^{2}}.
\label{hdmchd_11}
\end{eqnarray}

\subsection{Graphical display of results}

In two separate subsections below we present the results in the two 
scenarios, namely signal-dependent and signal-indepedent initiations. 
\subsubsection{Signal Dependent Initiation}

In Fig. (\ref{phase_1_dep})(a) and (b)  we present the phase diagrams 
of the system in $\alpha_1-\beta$ plane for $\gamma=0.9$ and $\gamma=0.1$, 
respectively. The reason for taking two extreme limits of $ \gamma$ for 
the two phase diagrams is to highlight how the phase boundaries get 
drastically affected by the variation of the hopping $W_{s}$ in the 
extended bottleneck region. Our theoretical predictions, based on MFA, 
are in good agreement with the corresponding numerical data  obtained 
from MC simulations for all phase boundaries except that between the 
HD/MC/LD and HD/MC/HD phases. 

The line separating the LD/LD/HD phase from HD/HD/HD phase is given by 
the expression (\ref{ldldhd_4}). The transition from LD/LD/HD phase 
to the HD/HD/HD phase is accompanied by a discontinues change in the 
coverage density of segment \RN{1} and \RN{2} while that of segment 
\RN{3} remains unaltered. Similarly, expression (\ref{hdmcld_12d}) 
represents the line that separates the HD/MC/HD phase from HD/MC/LD 
phase; during the transition across this line the coverage density of 
only the segment \RN{3} changes discontinuously. 


In Fig. (\ref{flux_3d}) we present 3D plots of the fluxes $J_{1}$ and 
$J_{2}$  as a function of (a) $\alpha_{1}$ and $\gamma$ at a fixed value of 
$\beta$, (b) $\gamma$ and $\beta$ at a fixed value of $s$, (c) $\alpha_{1}$ 
and $\beta$ for a fixed value of $\gamma$. One common feature of all the    
three 3D plots is that while one flux (say $ J_1$) increases then the other, 
namely $J_{2}$, decreases and the two surfaces cross each other. The 
suppression of $J_{2}$ by high value of the flux $J_{1}$ is displayed 
directly in the 2D cross sections of Fig. (\ref{flux_3d})(a) for different 
fixed values of $\gamma$ as shown in Fig.\ref{flux_2d_1}. 
{As $\gamma$ decreases, the areas on the $\alpha_{1}-\beta$ plane 
covered by  all the three phases, namely, LD/LD/MC, LD/LD/HD and 
HD/HD/HD, shrink.


The density profiles of the rods ($ \rho(i)=P_1(i)+P_2(i) $) are plotted 
in Fig. (\ref{density_profile}) for two sets of values of the parameters
$\beta, \alpha_{1}, \gamma$. The blue and red curves 
corresponds to the LD/LD/MC phase and LD/LD/HD phase, respectively. 
As is well known, the MFA is not as accurate for MC phase as it is 
for the LD and HD phases.

\subsubsection{Signal Independent Initiation}

The suppression of $J_{2}$ by $J_{1}$ is observed also in the case of 
signal-independent initiation just as the same phenomenon was observed 
in the case of signal-dependent initiation.Thus, this switch-like 
regulation of the fluxes of two mutually interfering species of rods 
seems to be an ubiquitous feature of the co-directional two-species 
exclusion process with two distinct sites of entry for the respective 
species  (see Fig.\ref{phase_1_ind} for phase diagram). 

The contour plots of fluxes and coverage densities in Figs.\ref{contour_plot_a} 
and \ref{contour_plot_b}, respectively, are consistent with the 
other results described above (see also Figs.\ref{flux_2d_3} and \ref{density_profile_2}).  

\section{Summary and conclusion}

We have developed a two-species exclusion process where each of the two species of hard rods 
have their distinct pair of entry and exit sites. An extended bottleneck exists in between the entry 
sites for the two species of rods; one entry site is located immediately downstream from the 
bottleneck while the other entry site is located far upstream from it. The static bottleneck is 
characterized by relatively slower rate of forward hopping of the rods in this region compared 
to that in the other regions. The number of species of the rods and the relative locations of their 
entry sites with respect to that of the bottleneck are motivated by an unconventional mode 
of gene translation via IRES. 

In principle, the results on the collective spatio-temporal organization 
of the rods in this exclusion model may have implications for regulation of unconventional 
translation. However, the work reported here is not intended to account for experimental data for 
any specific system. Instead, the  main emphasis has been to explore the generic features of the 
composite phases, fluxes and density profiles that are displayed by this simple model. More 
specifically we demonstrate how the boundaries between the various phases shift with the 
variation of some of the key model parameters that, at least in principle, can be varied in a 
controlled manner in experiments. But, out of all the seven possible phases, only a few may 
be realizable under physiological conditions unless a living cell has internal mechanism for 
altering the structure and/or stability of the secondary structure of the mRNA under different 
circumstances.

We have carried out our theoretical analysis in two distinct scenarios that correspond, 
respectively, to signal-dependent- and signal-independent IRES. Interestingly, in both 
scenarios we have observed switch-like regulation of the fluxes of the two species of rods: a 
sufficiently high-level of flux of one species can suppress that of the other to a low level. 
This phenomenon, that we propose to name as ``{\it translational interference}'', may be regarded 
as an analog of the well known phenomenon of ``transcriptional interference'' observed in the 
transcription of genetic message, encoded on DNA, by another type of motor called RNA polymerase.
However, there is a crucial difference between the two phenomena. Since mRNA track can form 
several types of secondary structures \cite{lewis17} these can create wide varieties of bottlenecks against 
translation whereas, to our knowledge, analogous bottlenecks are not created by any 
secondary structure of the DNA template in the transcriptional process. Therefore, an additional 
control is available in `translational interference' as compared to `transcriptional interference'.

\section{Acknowledgement}
We thank Kavita Jain and Tanweer Hussain for useful comments and suggestions.
This work has been supported by J.C. Bose National Fellowship (DC), ``Prof. S. Sampath Chair'' Professorship (DC) and by UGC Senior Research Fellowship (BM).

}
\appendix
\section{Calculation of $P(0, \dots,0)$}
\begin{figure}[t] 
\begin{center}
\includegraphics[width=1.0\columnwidth]{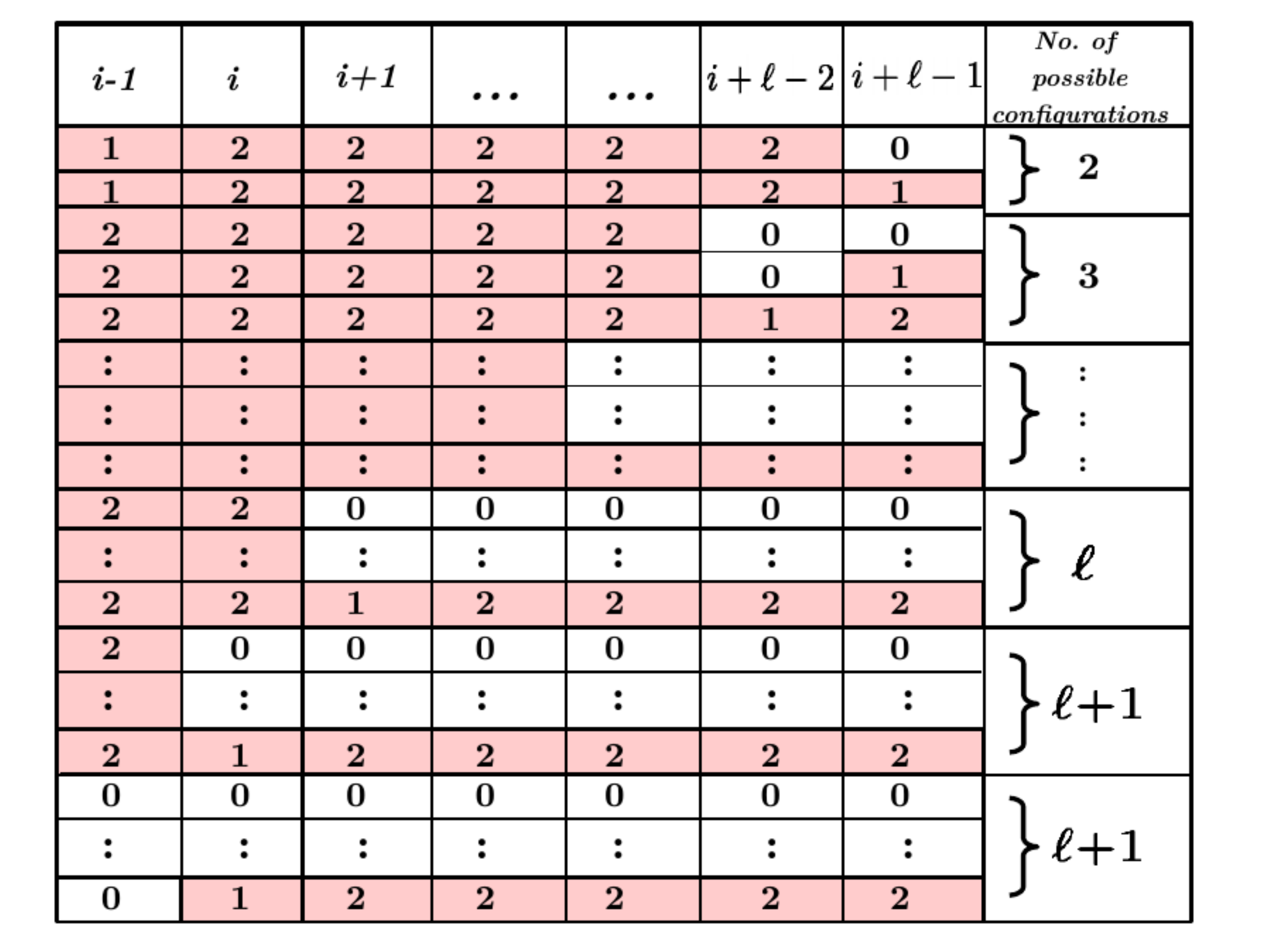} 
\end{center} 
\caption{Schematic diagram of all possible configurations of rods over $ \ell+1 $ consecutive sites. In their distribution we have followed total exclusion principle. Here, $ 0,1 $ and $ 2 $ represents the occupational status of any site. If it is $ ``0" $, then the site is neither occupied nor covered. If it is $ ``1" $, then it is occupied (i.e. the left most part of the rod is at this site) and if it is $ ``2" $, then it is covered by a rod.} 
\label{ires_model_n}
\end{figure}

We consider a system consisting of a long one dimensional lattice of size $ L+\ell-1 $ and 
multiple identical rods, each of size $ \ell>1 $.
We define a probability $ P(\underbrace{0, \dots,0}_{\ell})$ for the consecutive 
$ \ell $ sites being empty (i.e. neither covered nor occupied) simultaneously.
The total number $N$ of possible ways to distribute rods over $ \ell+1 $ consecutive sites 
(see Fig.\ref{ires_model_n}) is given by 
\begin{equation}
N=2+3+....+(\ell-1)+\ell+(\ell+1)+(\ell+1),
\label{number_1}
\end{equation}
\begin{equation}
N=\dfrac{(\ell-1)(\ell+2)+4(\ell+1)}{2}.
\label{number_2}
\end{equation}

Therefore, the probability that all consecutive $ \ell $ sites from site $ i$ to $ i+\ell-1 $ are not occupied is,\\
\begin{equation}
P(\underbrace{0, \dots,0}_{\ell}) = \dfrac{(1-\rho)^{\ell}}{\Theta},
\label{cond_prob_1}
\end{equation}
where,\\
\begin{eqnarray}
\Theta & =& \rho^2(1-\rho)^{\ell-1} + \rho(1-\rho)^{\ell} \\ \nonumber
&+ &  (\ell-2)\rho(1-\rho)^{\ell}  \\ \nonumber
&+ & \Bigg[\dfrac{\ell-2}{2} \Bigg]\rho^2(1-\rho)^{\ell-1}                                                                                                                                                                                                                                                                                                                                                                                                                                                                                                                                                                                                                                                                                                                                                                                                                                                                                                                                                                                                                                                                                                                                                                                                                                                                                                                                                                                                                                                                                                                                                                                                                                                                                                                                                                                                                                                                                                                                                                                                                                                                                                                                                                                                                                                                                                                                                                                                                                                                                                                                                                                                                                                                                                                                                                                                                                                                                                                                                 (\ell+1) \\ \nonumber
 &+ & (1-\rho)^{\ell+1} + \ell\rho(1-\rho)^{\ell}\\ \nonumber
 &+ & \rho(1-\rho)^{\ell} + \ell\rho^2(1-\rho)^{\ell-1},
 \label{theta_2}
\end{eqnarray}
is the sum of the weightage of all the possible configurations of the consecutive sites from $ i $ to $ i+\ell-1 $. 
On further simplification Eq. (\ref{cond_prob_1}) reduces to 
\begin{equation}
P(\underbrace{0, \dots,0}_{\ell}) = \dfrac{2(1-\rho)}{2(1-\rho)(1+\rho(\ell-1)) + \rho\ell(2+\rho(\ell-1))}.
\label{cond_prob_2}
\end{equation}
Next, we assume that inside each of the segments   \RN{1}, \RN{2} and \RN{3} the number density of the 
rods is uniform  and that it is $ \rho_{1}, \rho_{2} $ and $ \rho_{3} $ in the segments \RN{1}, \RN{2} and \RN{3},  respectively. Although this is a very good approximation in LD and HD phases, it is not so good in the MC phase.  
If special site is at $ i=i_s=L_2+n\ell+m $ and $ n>0 $, then,\\
\begin{equation}
P(\underbrace{0, \dots,0}_{\ell}) = \dfrac{2(1-\rho_{3})}{2(1-\rho_{3})(1+\rho_{3}^\prime) + \rho_3\ell(2+\rho_{3}^\prime)},
\label{cond_prob_3}
\end{equation}
where, $\rho_{3}^\prime=\rho_{3}(\ell-1)$.\\
\textbf{For LD phase in segment III}\\
\begin{eqnarray}
 \rho_{3}  =\alpha_{eff3} ,~~~~ 1-\rho_{3} =1-\alpha_{eff3},
 \label{ld_1}
\end{eqnarray}
\begin{equation}
P(\underbrace{0, \dots,0}_{\ell})=\dfrac{2(1-\alpha_{eff3})}{2(1-\alpha_{eff3})(1+\alpha_{eff3}^\prime) +\alpha_{eff3}\ell(2+\alpha_{eff3}^\prime)},
\label{ld_2}
\end{equation}
where, $ \alpha_{eff3}^\prime = \alpha_{eff3}(\ell-1) $.\\
\textbf{For HD phase in segment III}\\
\begin{eqnarray}
\rho_{3} =1-\beta,~~~ 1-\rho_{3} =\beta,
 \label{hd_1}
\end{eqnarray}
\begin{equation}
P(\underbrace{0, \dots,0}_{\ell})=\dfrac{2\beta}{2\beta(1+\beta^\prime) +(1-\beta)\ell(2+\beta^\prime)},
\label{hd_2}
\end{equation}
where, $\beta^\prime = (1-\beta)(\ell-1) $.\\
\textbf{For MC phase in segment III}\\
\begin{eqnarray}
\rho_{3} =\dfrac{1}{2},~~~~~ 1-\rho_{3}=\dfrac{1}{2},
 \label{mc_1}
\end{eqnarray}
\begin{equation}
P(\underbrace{0, \dots,0}_{\ell})=\dfrac{4}{2(\ell+1) +\ell(\ell+3)}.
\label{mc_2}
\end{equation}

\section{Phase 2}
\label{app-Phase2}

\begin{eqnarray}
\beta=&-&s_{0}-0.5 \sqrt{s_0-s_1-\frac{s_2}{s_3}-s_4} \\ \nonumber
&-&0.5 \sqrt{s_5-s_{6}+\frac{s_2}{s_3}+s_4-\dfrac{s_7}{s_{8}}}
\label{beta_1}
\end{eqnarray}
where
\begin{widetext}
\begin{eqnarray}
s_0=\frac{0.25 \left(-10\alpha_{eff2}^2-11\alpha_{eff2}^2-2\right)}{10 \alpha_{eff2}^2+1}
\label{s0}
\end{eqnarray}
\begin{eqnarray}
s_1=\frac{0.0009 \left(6400 \alpha_{eff2}^2-830 \alpha_{eff2}+493\right)}{10 \alpha_{eff2}+1}
\label{s1}
\end{eqnarray}
\begin{eqnarray}
s_2=0.005 \left(1.014\times 10^8 \alpha_{eff2}^4+6.32\times 10^7 \alpha_{eff2}^3-5.07\times 10^6 \alpha_{eff2}^2-241660 \alpha_{eff2}+1.09\times 10^6\right)
\label{s2}
\end{eqnarray}
\begin{eqnarray}
s_3=s_{31}(s_{32}+s_{33})^{1/3}
\label{s3}
\end{eqnarray}
where,\\
\begin{eqnarray}
s_{31}=10\alpha_{eff2}+1
\label{s31}
\end{eqnarray}
\begin{eqnarray}
s_{32}=&-&1.45\times 10^{15}\alpha_{eff2}^6-1.37\times 10^{15} \alpha_{eff2}^5-2.09\times 10^{14} \alpha_{eff2}^4+6.05\times 10^{13}\alpha_{eff2}^3 \\ \nonumber
&+&4.51\times 10^{13} \alpha_{eff2}^2+1.95\times 10^{12} \alpha_{eff2}-1.64\times 10^{12}
\label{s32}
\end{eqnarray}
\begin{eqnarray}
s_{33}=\sqrt{s_{331}+s_{332}+s_{333}}
\label{s33}
\end{eqnarray}
\begin{eqnarray}
s_{331}=-1.\times 10^{29} \alpha_{eff2}^{12}-1.47\times 10^{29} \alpha_{eff2}^{11}+2.48\times 10^{29}\alpha_{eff2}^{10}+2.95\times 10^{29} \alpha_{eff2}^9-1.92\times 10^{29} \alpha_{eff2}^8
\label{s331}
\end{eqnarray}
\begin{eqnarray}
s_{332}=-2.5\times 10^{29} \alpha_{eff2}^7-3.73\times 10^{28} \alpha_{eff2}^6+1.39\times 10^{28} \alpha_{eff2}^5+2.22\times 10^{27} \alpha_{eff2}^4-5.22\times 10^{26} \alpha_{eff2}^3
\label{s332}
\end{eqnarray}
\begin{eqnarray}
s_{333}=-1.059\times 10^{26} \alpha_{eff2}^2-4.56\times 10^{24} \alpha_{eff2}-8.69\times 10^{22}
\label{s333}
\end{eqnarray}
\begin{eqnarray}
s_4=s_{41}(s_{42}+s_{43})^{1/3}
\label{s4}
\end{eqnarray}
where,\\
\begin{eqnarray}
s_{41}=\frac{0.00004}{10\alpha_{eff2}+1}
\label{s4}
\end{eqnarray}
\begin{eqnarray}
s_{42}=s_{32}
\label{s4}
\end{eqnarray}
\begin{eqnarray}
s_{43}=s_{33}
\label{s4}
\end{eqnarray}
\begin{eqnarray}
s_5=\frac{0.5 \left(-10\alpha_{eff2}^2-11 \alpha_{eff2}^2-2\right)^2}{(10\alpha_{eff2}^2+1)^2}
 \label{s5}
\end{eqnarray}
\begin{eqnarray}
s_6=\frac{0.002 \left(6400 \alpha_{eff2}^2-830 \alpha_{eff2}+493\right)}{10\alpha_{eff2}+1}
\label{s6}
\end{eqnarray}
\begin{eqnarray}
s_7=0.025(s_{71}+s_{72}+s_{73})
\label{s7}
\end{eqnarray}
\begin{eqnarray}
s_{71}=-\frac{\left(-10\alpha_{eff2}^2-11\alpha_{eff2}-2\right)^3}{(10\alpha_{eff2}+1)^3}
\label{s71}
\end{eqnarray}
\begin{eqnarray}
s_{72}=-\frac{0.01 \left(2800\alpha_{eff2}^2-550\alpha_{eff2}+197\right)}{10\alpha_{eff2}+1}
\label{s72}
\end{eqnarray}
\begin{eqnarray}
s_{73}=\frac{0.01 \left(-10\alpha_{eff2}^2-11\alpha_{eff2}-2\right) \left(6400\alpha_{eff2}^2-830\alpha_{eff2}+493\right)}{(10 \alpha_{eff2}+1)^2}
\label{s73}
\end{eqnarray}
\begin{eqnarray}
s_8=\sqrt{s_0-s_{1}-\dfrac{s_2}{s_{3}}-s_{4}}
\label{s8}
\end{eqnarray}
\end{widetext}


\section{Phase 3}
\label{app-Phase3}
\begin{eqnarray}
\alpha_{eff2} = \frac{-t_4-\sqrt{t_4^2-4t_5t_6}}{2 t_6},
\label{a_eff2_1}
\end{eqnarray}
where,
\begin{eqnarray}
t_4=&-&8 \alpha_{IE}\ell^{3/2}-4 \alpha_{IE} \ell^2+8 \alpha_{IE} \sqrt{\ell}+4 \alpha_{IE} \nonumber \\
&-&10 \ell^{3/2} W_s-2 \ell^{5/2} W_s-\ell^3 W_s+\ell^3-6 \ell^2 W_s \nonumber \\
&+&4 \ell^2-7 \ell W_s-4 \sqrt{\ell} W_s-3 \ell-2 W_s-2,
 \label{t4}
\end{eqnarray}
\begin{eqnarray}
t_5&=&2 W_s - 
   4 \alpha_{IE} W_s - 8 \alpha_{IE}\sqrt{\ell} W_s + 5 \ell W_s  \nonumber \\
    &-& 4 \alpha_{IE} \ell W_s + \ell^2 W_s, 
\label{t5}
\end{eqnarray}
\begin{eqnarray}
t_6= 10 \ell^{3/2}+2 \ell^{5/2}+\ell^3+6 \ell^2+7 \ell+4 \sqrt{\ell}+2.
\label{t6}
\end{eqnarray}


\section{Phase 4}
\label{app-Phase4}
\begin{eqnarray}
\beta_{eff2}=\frac{-h_2-\sqrt{h_2^2-4h_1h_3}}{2h_1},
\label{b_eff2_1}
\end{eqnarray}
where,
\begin{eqnarray}
h_1 &=& \beta^3 \ell^3-4 \beta^3 \ell^2+5 \beta^3 \ell-2 \beta^3 
-2 \beta^2 \ell^3 \nonumber \\
&+&5 \beta^2 \ell^2 
-5 \beta^2 \ell+2 \beta^2 
+\beta \ell^3-2 \beta \ell^2 \nonumber \\
&+&\beta \ell+\ell^2+\ell, 
\label{h1}
\end{eqnarray}
\begin{eqnarray}
h_2&=&-2 \alpha_{IE} \beta^2 \ell^2+4 \alpha_{IE} \beta^2 \ell-2 \alpha_{IE}\beta^2-2 \alpha_{IE} \beta \ell  \nonumber \\
&+&2 \alpha_{IE} \beta+\beta^4 (-\ell^3)+4 \beta^4 \ell^2-5 \beta^4 \ell \nonumber \\
&+& 2 \beta^4-\beta^3 \ell^3 W_s+3 \beta^3 \ell^3+4 \beta^3 \ell^2 W_s-8 \beta^3 \ell^2 \nonumber \\
&-& 5 \beta^3 \ell W_s+7 \beta^3 \ell+2 \beta^3 W_s-2 \beta^3+2 \beta^2 \ell^3 W_s \nonumber \\
&-&\ 3 \beta^2 \ell^3-5 \beta^2 \ell^2 W_s+4 \beta^2 \ell^2+5 \beta^2 \ell Ws \nonumber \\
&-& \beta^2 \ell-2 \beta^2 W_s-\beta \ell^3 W_s+\beta \ell^3+2 \beta \ell^2 W_s \nonumber \\
&-& \beta \ell W_s-\beta \ell-\ell^2 W_s-\ell W_s,
\label{h2}
\end{eqnarray}
\begin{eqnarray}
h_3&=&-2 \alpha_{IE} \beta^2 \ell W_s+2 \alpha_{IE} \beta^2 W_s-2 \alpha_{IE} \beta W_s \nonumber \\
&+& \beta^4 (-\ell^2) W_s+3 \beta^4 \ell W_s 
-2 \beta^4 W_s \nonumber \\
&+& 3 \beta^3 \ell^2 W_s -5 \beta^3 \ell W_s +2 \beta^3 W_s-3 \beta^2 \ell^2 W_s \nonumber \\
&+&\beta^2 \ell W_s+\beta \ell^2 W_s+\beta \ell W_s.
\label{h3}
\end{eqnarray}
\\
\begin{align}
\beta_{eff1}=\frac{-q_1-\sqrt{q_1^2-4q_2q_3}}{2 q_3},
\label{b_eff1_1}
\end{align}
where,
\begin{eqnarray}
q_1&=&\beta_{eff2}^2 \ell-\beta_{eff2}^2-\beta_{eff2} \ell W_s+\beta_{eff2} \ell  \nonumber \\
&+&\beta_{eff2} W_s-\beta_{eff2}+W_s,
\label{q1_1}
\end{eqnarray}
\begin{eqnarray}
q_2=\beta_{eff2}^2-\beta_{eff2} W_s,
\label{q2_1}
\end{eqnarray}
\begin{eqnarray}
q_3=\beta_{eff2}(-\ell)+\beta_{eff2}-W_s.
\label{q3_1}
\end{eqnarray}

\section{Phase 7}
\label{app-Phase7}

\begin{eqnarray}
\beta=z_0-0.5 \sqrt{z_1+\frac{z_2}{z_3}+z_4} 
-0.5 \sqrt{z_5-\frac{z_6}{z_7}-z_8-\frac{z_9}{z_{10}}} \nonumber \\
\label{hdmchd_9}
\end{eqnarray}
where,\\
\begin{eqnarray}
z_0=0.75
\label{z0}
\end{eqnarray}
\begin{widetext}
\begin{eqnarray}
z_1= 5.5 \times 10^{-11} (-5.2 \times 10^{10} - 4.5 \times 10^9 \alpha_{IE}) 
 + 2.2 +1.8 \times 10^{-11} (5.2 \times 10^{10} + 4.5 \times 10^9 \alpha_{IE}) \nonumber \\
 \label{z1}
\end{eqnarray}
\begin{eqnarray}
z_2= 2.310^{-11} (2.0\times 10^{19} \alpha_{IE}^2 +5.4\times 10^{20} \alpha_{IE} 
+2.7\times 10^{20})
\label{z2}
\end{eqnarray}
\begin{eqnarray}
z_3=\sqrt[3]{z_3^+  
- \sqrt{z_3^-}}
\label{z3}
\end{eqnarray}
where,\\
\begin{eqnarray}
z_3^+=1.7\times 10^{29} \alpha_{IE}^3+7.4\times 10^{30}\alpha_{IE}^2 +3.2\times 10^{31} \alpha_{IE} +8.5\times 10^{30}
\label{z31}
\end{eqnarray}
\begin{eqnarray}
z_3^-=4.2\times 10^{58} \alpha_{IE}^5-5.9\times 10^{60} \alpha_{IE}^4-2.4\times 10^{62} \alpha_{IE}^3+1.8\times 10^{62} \alpha_{IE}^2+7.1\times 10^{61} \alpha_{IE}-6.4\times 10^{60}
\label{z32}
\end{eqnarray}

\begin{eqnarray}
z_4=1.5 \times 10^{-11} z_3
\label{z4}
\end{eqnarray}
\begin{eqnarray}
z_5=4.6 - 7.4\times 10^{-11} (5.2\times 10^{10} +4.5\times 10^{9}\alpha_{IE})
\label{z5}
\end{eqnarray}
\begin{eqnarray}
z_6=2.3\times 10^{-11} (2.0\times 10^{19} \alpha_{IE}^2 +5.4\times 10^{20} \alpha_{IE} 
+2.7 \times 10^{20})
\label{z6}
\end{eqnarray}
\begin{eqnarray}
z_7=z_3
\label{z7}
\end{eqnarray}
\begin{eqnarray}
z_8=1.5\times 10^{-11}z_7
\label{z8}
\end{eqnarray}
\begin{eqnarray}
z_9=0.25 (-4.5\times 10^{-10} (4.9\times 10^8 \alpha_{IE}-1.7\times 10^{10}) 
-6.7^{-10} (4.5\times 10^9 \alpha_{IE}+5.2\times 10^{10})+27.9)
\label{z9}
\end{eqnarray}
\begin{eqnarray}
z_{10}=z_1+\frac{z_2}{z_3} + z_4
\label{z10}
\end{eqnarray}

\end{widetext}

\end{document}